\documentclass[reprint,aps,amsmath,amssymb]{revtex4-2}

\usepackage{graphicx}
\usepackage{dcolumn}
\usepackage{bm}
\usepackage{algpseudocode}
\usepackage[dvipsnames, table]{xcolor}
\usepackage{comment}

\begin{document}

\title{Physics-Guided Actor-Critic Reinforcement Learning for Swimming in Turbulence}
\author{Christopher Koh}
\author{Laurent Pagnier}
 \email{laurentpagnier@arizona.edu}
\author{Michael (Misha) Chertkov}
 \email{chertkov@arizona.edu}
\affiliation{
 Applied Mathematics \& Mathematics, University of Arizona, Tucson, AZ 85721.
}

\date{\today}

\begin{abstract}
Turbulent diffusion causes particles placed in proximity to separate. We investigate the required swimming efforts to maintain an active particle close to its passively advected counterpart. We explore optimally balancing these efforts by developing a novel physics-informed reinforcement learning strategy and comparing it with prescribed control and physics-agnostic reinforcement learning strategies. Our scheme, coined the actor-physicist, is an adaptation of the actor-critic algorithm in which the neural network parameterized critic is replaced with an analytically derived physical heuristic function, the physicist. We validate the proposed physics-informed reinforcement learning approach through extensive numerical experiments in both synthetic BK and more realistic Arnold-Beltrami-Childress flow environments, demonstrating its superiority in controlling particle dynamics when compared to standard reinforcement learning methods.
\end{abstract}

\maketitle

\section{Introduction}

\begin{figure}[ht]
    \centering
    \includegraphics[width=\linewidth]{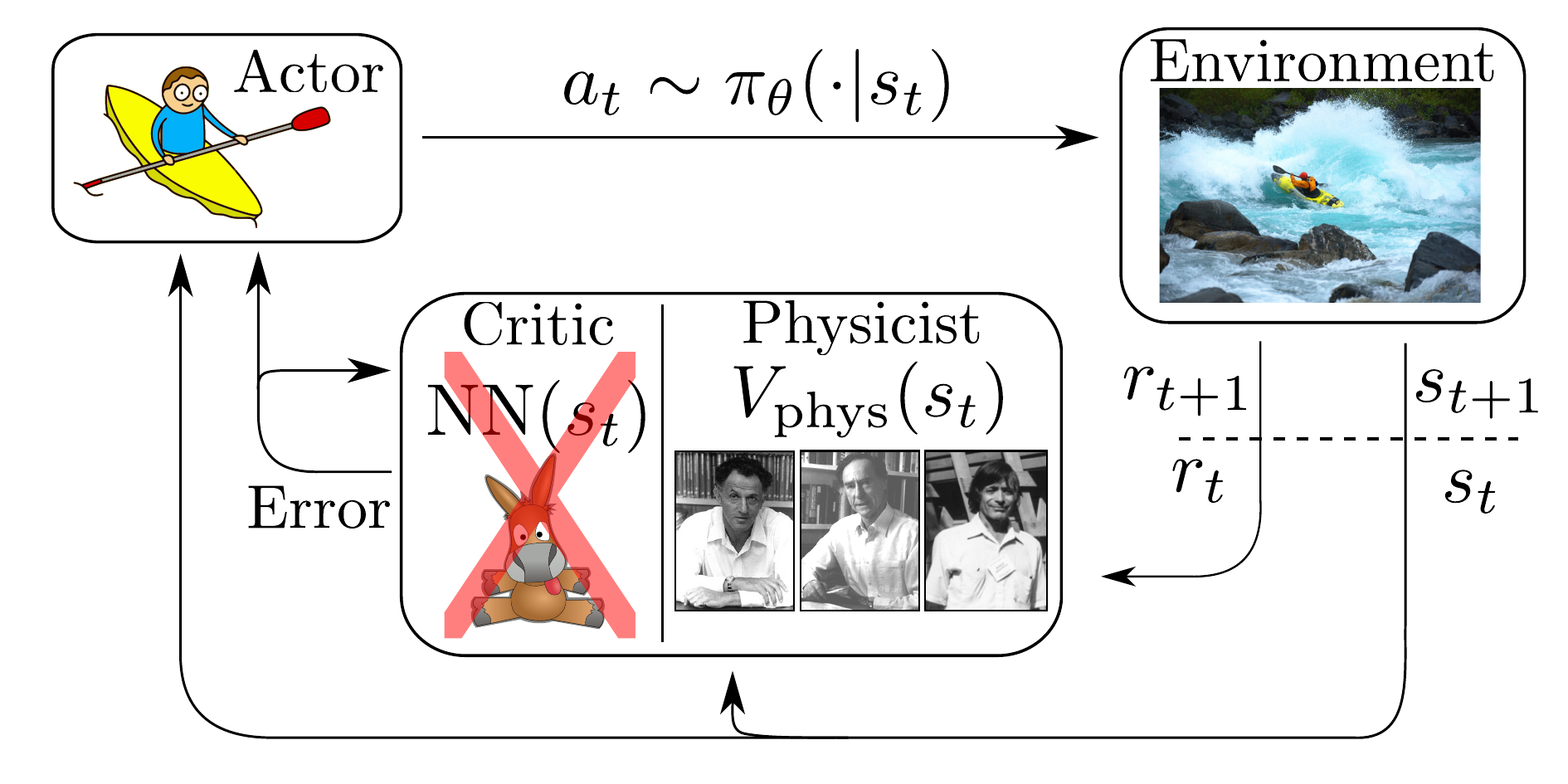} 
    \caption{Actor-Physicist (AP) diagram. The main idea of the present works is to substitute an expression obtained from the system's dynamics in place of the standard neural network.}
    \label{fig:ac_cartoon}
\end{figure}

This article's key technical advancement is the development of a Physics-Informed Reinforcement Learning (PIRL) approach where a \textit{physicist} replaces the \textit{critic} in the standard Actor-Critic (AC) algorithm, see Fig.~\ref{fig:ac_cartoon}. We coin it Actor-Physicist (AP) algorithm. The physicist component leverages physical insights on the statistics and control of swimming in chaotic flows, as elaborated in \cite{chertkov_universality_2023} and further developed in this manuscript. This approach, derived from our understanding of Lagrangian separation in turbulent flows, guides the control policies.

We consider a particle in a turbulent flow that swims towards its passive target to maintain proximity. The particle will be controlled according to Reinforcement Learning (RL) \cite{sutton_reinforcement_2018} -- a methodology for solving complex decision-making problems. RL involves an agent learning through interaction with its environment, balancing exploration and exploitation. Exploration involves trying new actions to gain information about the turbulent environment while exploitation uses accumulated knowledge to make optimal decisions. Further details on the RL framework are provided in Section~\ref{sec:RL}, which introduces this topic  using physics-friendly terminology. This RL decision-making is linked to Stochastic Optimal Control (SOC), where the agent maximizes expected reward under environmental uncertainty. Our reward function in the SOC for particles consists of two competing terms: one maintaining distance between the agent (i.e., the active particle) and its target (passive particle advected by the flow) and the other term penalizing the effort required. 

\begin{figure}[ht]
    \centering
    \includegraphics[width=\linewidth]{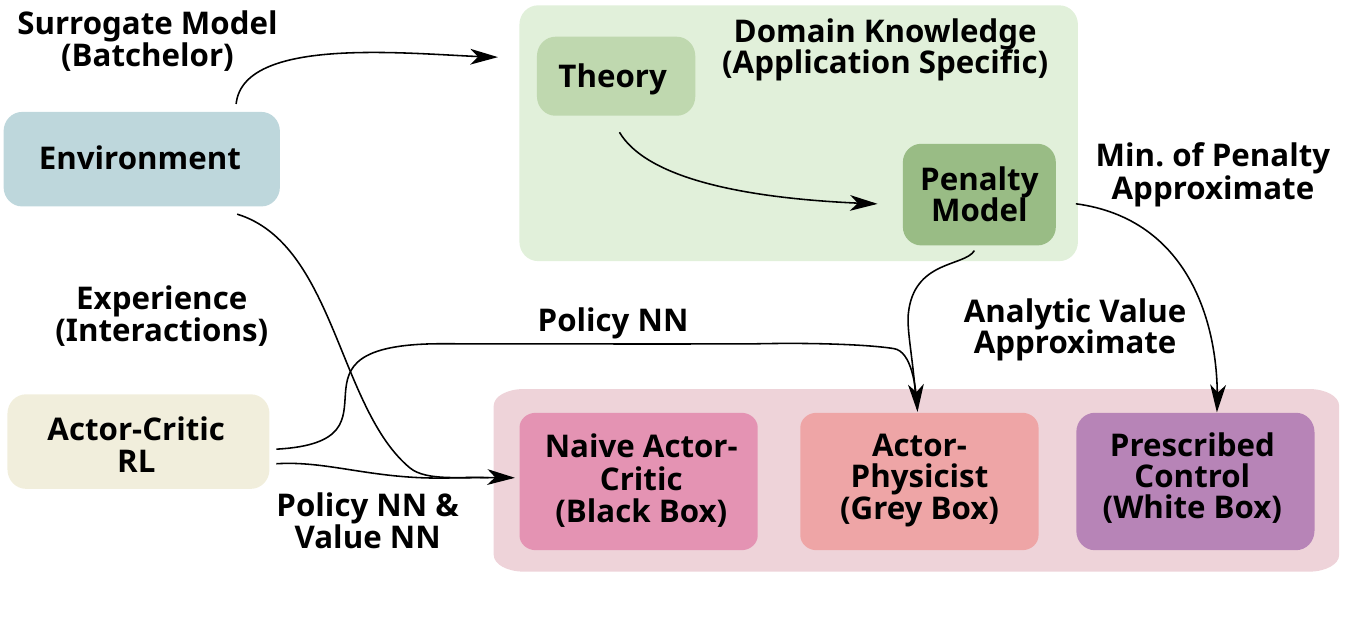} 
    \caption{Flowchart explaining the relations between RL, environment, theory, and the components of the different control schemes.}
    \label{fig:flowchart}
\end{figure}

It is important to emphasize that we arrive at the RL solution in a number of steps, which included analytical analysis reported earlier in \cite{chertkov_universality_2023}, and which we briefly review in Section \ref{sec:setup}, and also extended in this manuscript. The main handicap of the analytic analysis is that it deals with an idealized setting where only a prescribed -- and thus only conservative (suboptimal) control applies. Therefore, RL becomes the universal and data-driven tool which allows us to overcome the handicap of the theory and help to develop a much more general, robust, data-driven and assymptotically optimal approach to controlling the separation between active and passive particles.  Additional nuances of this manuscript RL approach are revealed in the flowchart of relations between different methods applied to the problem of the swimming control in realistic flows illustrated in Fig.~(\ref{fig:flowchart}) which we briefly outline next. 

\underline{Swim Control in Realistic Flows:} 
Our primary objective is to maintain proximity between passive and active particles by maximizing a time-integrated and averaged reward function. We develop and evaluate several control policies for the active particle, tailored to the available information about flow statistics (leading Lyapunov exponent):
\begin{itemize}
\item {\it Prescribed Control:} In this case, used as a benchmark for the RL schemes described in the following, the control force exerted by the active particle is proportional to the separation with the coefficient of the proportionality prescribed/fixed.

\item {\it Actor-Critic Agents:} This is a standard RL architecture. Here, both the policy (actor) and a guiding principle based on the so-called value function (critique) are built based on Neural Networks (NN).  In this work, we use two AC methods for comparison with our AP method: Advantage Actor-Critic (A2C) \cite{mnih_asynchronous_2016} and Proximal Policy Optimization (PPO) \cite{schulman_proximal_2017}.

\item {\it Actor-Physicist Agent:} This method modifies the AC framework, in this work the actor side of our AP agent is based on the A2C architecture \cite{mnih_asynchronous_2016}. While the physicist side -- replacing critic -- uses an analytically derived prescribed control estimate of the value function instead of its NN-version. This analytic estimation of the value function -- based on the Stochastic Optimal Control (SOC) formulation applied to an idelialized, so-called Batchelor-Kraichnan (BK) modeling of the stochastic flow, but applied to a realistic flow -- is described in Section \ref{sec:value-SK}. 
\end{itemize}

Fig.~\ref{fig:flowchart} summarizes the relations between these methods and the environment and the theory.  On the left of the chart, we see two fundamental notions of the Reinforcement Learning -- the environment and the agent. The upper block illustrates that the theory provides a base for the models and approaches discussed. In the lower block of the chart, we list all the models we discuss and use in the manuscript. Starting to the right, one has the most interpretable control which rely on strong assumptions that may not be satisfied by the environment. Moving to the left, control becomes less interpretable. Arrows describe relations between the models and how they acquire information about the environment (physics of the stochastic flows).

\subsection*{Relevant Approaches by Others}

Optimally navigating an agent through a fluid is a challenging and active research topic, with recent reinforcement learning approaches for controlling particles and related systems summarized in \cite{garnier_review_2021} and \cite{rabault_deep_2020}. Single-agent RL methods were developed for controlling particle (or solid body) navigation specific for efficient gliding in natural stratified air-flow environments \cite{reddy_glider_2018, novati_controlled_2019} and for autonomous navigation along an optimal point-to-point path in various complex fluid flows \cite{biferale_zermelos_2019, alageshan_machine_2020, gunnarson_learning_2021}. RL with multiple, or at least two, conflicting objectives was also extended to a competitive game played by two active particles in a turbulent environment, with one attempting to catch and the other aiming to escape \cite{borra_reinforcement_2022}. In a setting closely related to the one discussed in this manuscript, the authors of \cite{calascibetta_taming_2023} devised a multi-objective (pareto-seeking) RL approach aimed at controlling separation and minimizing efforts between two active particles by switching in time between a number of prescribed strategies, with observations of the separation and velocity difference available at all times. Notably, \cite{biferale_zermelos_2019} utilized the Actor-Critic (AC) version of RL, which is central to this manuscript, although their critic is modeled via a neural network in a physics-agnostic (of flow) manner. The RL study in \cite{calascibetta_taming_2023} was complemented by a parallel study by the same authors \cite{calascibetta_optimal_2023}, which juxtaposed a stochastic optimal control approach to the problem of an active particle following a passive particle in a turbulent flow with a heuristic strategy that relies on the particle's local observation of the environment. 

Most of the RL papers mentioned above, with the exceptions of \cite{borra_optimal_2021, borra_reinforcement_2022, calascibetta_taming_2023}, utilize RL in a domain-agnostic, physics-uninformed manner, relying heavily on feature engineering of the underlying neural networks to achieve satisfactory performance. In contrast, in this manuscript we utilize domain knowledge through value approximation and do not require enhancements in the modeling of the agents themselves. Additionally, our physics-informed RL approach helps address the difficulties traditional AC methods face in learning values across multiple orders of magnitude \cite{acrossvalues} and serves as an alternative to jump-start or imitation learning \cite{uchendu2023jumpstart} when learning from a warm start. This makes our principal physics-guided/informed approach relevant to many other navigation/control challenges beyond the context of RL for swimming where physical modeling of environment, even if approximate, is available. 

\section{Reinforcement Learning} \label{sec:RL}

The main part of this section provides a brief technical introduction aimed at readers with a physics background but without prior experience in Reinforcement Learning (RL) and its Actor-Critic (AC) framework (see, e.g., \cite{sutton_reinforcement_2018}). Readers already familiar with RL may wish to skip this part and proceed directly to Subsection \ref{sec:PIRL}, where we introduce our custom and physics-informed modification of the AC consisted in replacement of the "standard" critic by a physicist-critic -- a critic guided by a physical model of the environment.

RL is a data-driven version of Stochastic Optimal Control (SOC) which in its continuous time formulation with finite time horizon (episode) can be stated as follows:
\begin{align}\label{eq:SOC}
\max_{\pi}  \int_{0}^T dt e^{-\nu t}\,\mathbb{E}_{a\sim\pi}\Big[r(s(t),a(t))\Big], 
\end{align}
where the $s(t)$  and $a(t)$ are  state observed and action taken according to the policy $\pi$ at the moment of time $t$, $\nu$ is the discount factor that tells the agent whether to prioritize short-term or long-term rewards, and $r(s(t),a(t))$ is the reward encouraging or penalizing actions of the agents.

In RL, which should be considered as a data driven version of SOC, an agent interacts with an environment, which evolves in discrete time steps. Let us now explain  SOC (\ref{eq:SOC}) on its discrete time RL version. Both in SOC and RL the environment is uncertain, and the next state $s'$ is sampled based on the current state $s$ and the agent's action $a$: $s' \sim p(\cdot|s,a)$, where the probability distribution is unknown (or at least not fully known). Then Eq.~(\ref{eq:SOC}) turns to  
\begin{equation}
    \mathcal{G}_\pi(T) = \Delta \mathbb{E}_{\substack{s_{0\to N}\\a_{0\to N}}}\Big[\sum_{n=0}^N\gamma^{n} r(s_{n},a_{n})\Big], \label{eq:avg_return}
\end{equation}
where $\pi(a_n|s_n)$ is the policy which defines the probability for the agent to select an action $a_n = a(t_n)$ conditioned to the current state $s_n = s(t_n)$. Here, $t_n = n \Delta = n T/N$, $n = 0, \cdots, N \to \infty$, and $0 < \gamma \leq 1$ is the discrete time version of the discount factor $\nu$ in Eq.~(\ref{eq:SOC}). In Eq.~(\ref{eq:avg_return}), $s_{n\to N}$ and $a_{n\to N}$ are shorthand notations for $(s_{k+1} \sim p(\cdot|s_k, a_k)| k=n, \cdots, N-1)$ and $(a_k \sim \pi(\cdot|s_k)| k=n, \cdots, N-1)$, respectively.

It is also custom in the field to define the state-action and state value functions:
\begin{align} \label{eq:Q}
Q_\pi(s_n,a_n) &= \Delta \mathbb{E}_{\substack{s_{n\to N}\\a_{n\to N}}}\Big[\sum_{k=n}^N\gamma^{k-n} r(s_k,a_k)\, \big|\, s_n, a_n\Big], \\
V_\pi(s_n) &=  \mathbb{E}_{a_n\sim \pi(\cdot|s_n)}\Big[ Q_\pi(s_n,a_n)\Big], \label{eq:V}
\end{align}
which describe the expected reward accumulated from the current time $t_n = n T/N$ until the end of the episode, under the given policy and conditioned on the current state and action ($s_n$ and $a_n$ in Eq.~(\ref{eq:Q})) and the current state ($s_n$ in Eq.~(\ref{eq:V})).

A classic RL approach derives recursive equations for the value functions under a given policy, followed by policy optimization in a greedy, dynamic programming manner. In this manuscript, we adopt a policy gradient-based method that focuses on maximizing the value function over the policy. In the age of AI, the policy function is parameterized via a Neural Network (NN) as $\pi_\theta(\cdot)$, where $\theta$ represents the parameters of the NN. The objective is to find the maximum of $V_{\pi_\theta}(s)$ by evaluating its gradient with respect to $\theta$ and iteratively updating $\theta$ to make the gradient zero in the limit:
\begin{equation}
\mathbb{E}_{a\sim \pi_\theta(\sim|s)}\left[Q_\pi(s,a)\nabla_\theta \log \pi_\theta(a|s)\right] = 0. \label{eq:pol_grad_th}
\end{equation}
However, this method tends to have high variance, resulting in slow convergence and unreliable estimates. To address this issue, it was suggested to replace the state-action value function in Eq.~(\ref{eq:pol_grad_th}) with the so-called \textit{advantage function} \cite{sutton_reinforcement_2018}:
\begin{equation*}
A_\pi(s,a) = Q_\pi(s,a) - b(s), 
\end{equation*}
where the \textit{baseline} $b(s)$ depends only on the state and not on the action. This ensures that:
\begin{equation}
\mathbb{E}_{a\sim \pi_\theta(\sim|s)}\left[A_\pi(s,a)\nabla_\theta \log \pi_\theta(a|s)\right] = 0, \label{eq:pol_grad_th-A}
\end{equation}
assuming Eq.~(\ref{eq:pol_grad_th}) is valid, 
due to the fact that $\mathbb{E}_{a\sim \pi_\theta(\sim|s)}\left[\nabla_\theta \log \pi_\theta(a|s)\right] = 0$ for any $s$. Although $b(s)$ can be any function of $s$, a common choice is $b(s) = V_\pi(s)$.

This modification -- from $Q$ to $A$ with $b=V$ -- results in the widely-used \textit{actor-critic} methods, where in addition to the policy function (the \textit{actor}), the state-value function $V(\cdot)$ (the \textit{critic}) is also approximated by a NN.

This actor-critic modification of the vanilla policy gradient (where only an actor is present, without a critic) raises the question: how should we interpret this baseline? Formally, it is a degree of freedom used to reduce the variance in the value function's gradient. Informally, the baseline serves as a benchmark for comparison.

Though the use of $V_\pi$ as a baseline is powerful, it has its limitations. The main drawback is that the training objective changes significantly with each NN update, resulting in continued variations in the gradient estimates. While there are partial solutions, such as delaying updates to the baseline \cite{mnih2013playing}, we propose a physics-informed remedy.

\subsection{Actor-Critic Reinforcement Learning with Physics Informed Critic}\label{sec:PIRL}

In this manuscript, we elevate the concept of the baseline/critic/benchmark by using a physically derived estimate for the state-dependent value function, rather than relying on a NN as in standard physics-agnostic actor-critic RL approaches like \cite{mnih2013playing}. In the following sections, we demonstrate that under certain simplifying assumptions about system dynamics and control, an explicit analytical expression for the baseline, as a function of the state, can be derived for a pair of particles placed in a large-scale turbulent flow.

\begin{table}[h!]
\centering
\begin{tabular}{ ll }
\hline
 \multicolumn{1}{c}{\vphantom{$\sum_f^f$}\textbf{General RL}} &  \multicolumn{1}{c}{\textbf{PIRL for Particles}} \\
\hline
Agent & Active particle \vspace{4pt}\\ 
State, $s$ & Separation between active and\vspace{4pt}\\
& passive particles, ${\bm s}$ \vspace{4pt}\\ 
Action, $a \sim \pi(\cdot|s)$ & Swimming efforts of the active swim- \\
 & mer\vspace{2pt}\\
Transition proba-\hphantom{ility} & Stochastic evolution according to rel- \\
ility $p(s'\,|\,s,\, a)$& ative turbulent velocity between swim-\\
 &   mers and Brownian force, as discussed \\
&   in Section \ref{sec:swimming-intro}.\vspace{4pt}\\
Baseline, $b(s)$ & State value function, $V_\phi({\bm s})$, estimated \\
& following assumptions of the BK the- \\
& ory, as discussed in Section \ref{sec:opt-stat-BK}.\\
\hline 
\end{tabular}
\caption{Mapping of general RL notations to the specific problem of controlling the relative separation of two particles.}
\label{tab:map}
\end{table}

Table \ref{tab:map} presents the mapping of general RL notations to the specific problem of controlling the relative separation between two particles. In this context, the agent is represented by the active particle, whose state is the separation distance from the passive particle. The action corresponds to the swimming efforts of the active particle, while the transition probability models the stochastic evolution influenced by relative velocity and Brownian forces.

We also present below a joint pseudo-algorithm for the Actor-Critic (AC) and Actor Physicist (AP) where the difference between the two is highlighted: 
\begin{algorithmic}
    \Require Number of episodes $N$. Each episode is split into $T$ intervals. Learning rates $\alpha, \beta >0$. Discount factor $\gamma \in [0,1]$.
    \State initialize NN for the policy function $\pi_\theta$ ($\theta$ is the NN's vector of parameters)
    \State \textcolor{red}{AC: initialize NN for the value function $V_w$}
    \State\textcolor{blue}{AP: select $\phi$ for the value function $V_\phi$}
    \State Initialize $s_0$
    \For{$episode = 1$ \textbf{to} $N$}
        \For{$t=1$ \textbf{to} $T$} 
            \State $a_t \sim \pi_\theta(\cdot | s_t)$
            \State Take action $a_t$, observe $s_{t+1}$, $r_t$
            \State  \textcolor{red}{AC: $A_t \gets r_t + \gamma V_w(s_{t+1}) - V_w(s_{t})$}
            \State  \textcolor{blue}{AP: $A_t \gets r_t + \gamma V_\phi(s_{t+1}) - V_\phi(s_{t})$}
            \State $\theta \gets \theta + \frac{\alpha}{t}\sum_{i=0}^{t} \nabla_{\theta} \log(\pi_{\theta}(a_t|s_t))A_t$    
        \EndFor
    \EndFor
\end{algorithmic}

\section{Preliminaries: Prescribed and Optimal Control of Swimming in Turbulence} \label{sec:setup}

In this introductory Section we follow \cite{chertkov_universality_2023} where stochastic flows, as well as a steady state stochastic optimal control, were discussed in an idealized setting. The goal of this section is to set stages for further discussion which is original to this manuscript going beyond the preliminary theoretical estimations of \cite{chertkov_universality_2023} and using these to develop general RL approach for dealing with a realistic flow, discussed in the following. 

We start this section in Section \ref{sec:swimming-intro} by setting up the basic equations for the stochastic dynamics and the statistics of separation between pairs of particles in a general large scale chaotic flow conditioned on a prescribed control. These expressions will depend on the finite-time statistics of the largest Lyapunov exponent of the chaotic flow, which is also discussed here. 

Then in Section \ref{sec:steady-general} we will describe how the general theory is extended to the case of the prescribed linear control. It will allow us to get expression for the probability distribution of inter-swimming separation. We observe that even in  the case of sufficiently strong prescribed control the probability distribution shows extended algebraic tails. However, the prescribed control is too conservative, which leads us to the Stochastic Optimal Control (SOC) formulation  in Section \ref{sec:opt-stat-BK} where we briefly discuss how the SOC optimal value of the linear control can be found in the stationary regime of the SK flow.

\subsection{Two Particles in a Chaotic Flow}\label{sec:swimming-intro}

We consider a spatially smooth, large-scale chaotic velocity field with a zero mean velocity. This type of flow has been extensively studied in stochastic hydrodynamics, see \cite{batchelor_small-scale_1959, kraichnan_small-scale_1968, shraiman_lagrangian_1994, chertkov_statistics_1995, bernard_slow_1998, balkovsky_universal_1999}, and the relevant review \cite{falkovich_particles_2001}. 
Particles placed in such flows are assumed to be very small relative to the typical length scales of the flow and they thus do not affect or alter the flow itself, making interactions between particles negligible. The particles separate exponentially fast. Our task is to navigate an active particle to control its separation from a passive target, assuming they were released in almost the same position initially.  Let ${\bm s}_{\alpha}(t) = (s_{\alpha;i}\,|\,i=1,\cdots,d)$ represent the positions of the two particles, $\alpha=1,2$, in $d$-dimensional flow (where $d=2$ or $d=3$). The separation vector ${\bm s} = {\bm s}_{1} - {\bm s}_{2}$ evolves according to:
\begin{gather}\label{eq:swim-r}
    \frac{{\rm d}\bm s}{{\rm d}t} - {\bm v}(t,\,{\bm s}_{1},\,{\bm s}_{2}) = -\frac{{\bm a}(t) + {\bm \xi}(t)}{\tau},
\end{gather}
where $\tau$ is the friction coefficient (set to unity for simplicity), ${\bm a}(t) = (a_i(t)|i=1,\cdots,d)$ is the control term that represents the active particle's action to fight against the flow, and ${\bm \xi}(t)$ is the difference in Brownian forces acting on the active and passive particles, modeled as zero-mean white-Gaussian noise:
\begin{gather*}
\forall i,j:\quad \mathbb{E}\big[\xi_i(t)\,\xi_j(t')\big] = \kappa\, \delta_{ij}\, \delta(t-t'),
\end{gather*}
where $\kappa$ is the diffusion coefficient.

The relative velocity vector ${\bm v}(t, {\bm s}_{1}, {\bm s}_{2}) := {\bm v}_{1}(t, {\bm s}_{1}) - {\bm v}_{2}(t, {\bm s}_{2})$ will be modeled differently throughout the manuscript. Following standard stochastic hydrodynamics assumptions \cite{falkovich_particles_2001} for large-scale flows, we approximate ${\bm v}(t, {\bm s}_{1}, {\bm s}_{2})$ by the leading term in its Taylor expansion in ${\bm s}$:
\begin{gather*}
{\bm v}(t,\, {\bm s}_{1},\, {\bm s}_{2}) \approx {\bm \sigma}(t) \,{\bm s},
\end{gather*}
where ${\bm \sigma}(t)$ is a possibly time-dependent velocity gradient matrix.  For the \textit{general} case, we consider an auxiliary multiplicative dynamics:
\begin{gather}\label{eq:W}
    \frac{\rm d}{{\rm d}t'}{\bm W}(t';t) = {\bm \sigma}(t')\, {\bm W}(t';t),
\end{gather}
where ${\bm W}(t';t) \in \mathbb{R}^{d \times d}$, with ${\bm W}(t;t) = \mathbb{I}_{d\times d}$, representing the time-ordered exponential of ${\bm \sigma}(t)$, denoted as ${\bm W}(t';t) = \mathcal{T}\big[\exp\big(\int_{t}^{t'} {\rm d}t'' {\bm \sigma}(t'')\big)\big]$. 

According to the Oseledets theorem \cite{ruelle_ergodic_1979, goldhirsch_stability_1987, falkovich_particles_2001}, at sufficiently large times, the matrix $\log ({\bm W}^\top(t';t){\bm W}(t';t))$ $/\,(t'-t)$ stabilizes, with eigenvectors tending to $d$ fixed orthonormal eigenvectors, ${\bm f}_i$, and the eigenvalues $\lambda_i(t';t) = \log |{\bm W}(t';t) {\bm f}_i|\,/\,(t'-t)$ stabilizing to their mean values. The finite-time statistics of $\lambda_i$ is given by:
\begin{align}\label{eq:Cramer}
& P\big(\lambda_1(t';t), \cdots, \lambda_d(t';t)\,|\,t'-t\big)  \\
  & \!\! \propto \exp\Big[-(t'-t)\, S\big(\lambda_1(t';t), \cdots, \lambda_d(t';t)\big)\Big], \nonumber 
\end{align}
where $S(\cdot)$, called the Cram\'er function, is convex, and the Lyapunov exponents are ordered: $\lambda_1 \geq \lambda_2 \geq \cdots \geq \lambda_d$. Of particular interest are the finite-time statistics of the leading Lyapunov exponent:
\begin{gather}\label{eq:Cramer-1}
    P(\lambda_1(t';t)\,|\,t'-t) \propto \exp\Big[-(t'-t)\, S_1\big(\lambda_1(t';t)\big)\Big].
\end{gather}

This manuscript focuses primarily on the reduced description of the flow associated with the statistics of the relative separation of two particles, allowing us to limit our discussion to the finite-time statistics of $\lambda_1(t';t)$. Computing $S_1(\cdot)$ -- the Cram\'{e}r function of $\lambda_1(t';t)$ -- analytically is possible only for a limited number of special, idealized flows, such as the stochastic Batchelor-Kraichnan (BK) flow \cite{chertkov_exact_1994, chertkov_statistics_1995, balkovsky_universal_1999}, in which case the Cram\'{e}r function is positive, quadratic. Therefore, our approach to computing $S_1(\lambda_1)$ in large-scale chaotic flows -- such as discussed in Section \ref{sec:ABC} for our running example of the ABC flow -- will be empirical. We extract it from simulations by placing two particles close to each other initially, tracking their separation over time, and then computing the Cram\'{e}r function by building statistics of the log-separation divided by time, accumulated over many trials, and fitting it to the expression on the right-hand side of Eq.~(\ref{eq:Cramer-1}).

\subsection{Batchelor Flow under Proportional Control}\label{sec:steady-general}
Consider an active particle moving towards its passive counterpart governed by a linear-feedback and time-independent force. This controlled motion can be expressed as
\begin{equation*}
{\bm a}\to \phi \, {\bm s}.
\end{equation*}
Here, $\phi$ is a constant (time-independent) parameter. Assuming that $\phi$ is sufficiently large, we focus our analysis on the steady state of the probability distribution function representing distance between the two particles. We discuss this scenario under two contexts: first in the general Batchelor case and then in the special short-time correlated case of the BK model.

In the general Batchelor  case solution of Eq.~(\ref{eq:swim-r}) for ${\bm s}(t)$ observed at the time $t$ conditioned to ${\bm s}(0)={\bm s}_0$, becomes
\begin{align}\label{eq:swim-solution}
    {\bm s}(t)=e^{-\phi t}\,{\bm W}(t)\bigg({\bm s}_0\!+\!\int_0^t \!{\rm d}t' e^{\phi t'} {\bm W}^{-1}(t')\,{\bm \xi}(t')\bigg),
\end{align}
where the time-ordered exponential, ${\bm W}(t)$, satisfies Eq.~(\ref{eq:W}). We will study the large time statistics of $s(t)=|{\bm s}(t)|$, where the initial separation is forgotten, and thus the first term within the brackets on the right hand side of Eq.~(\ref{eq:swim-solution}) can be dropped. Moreover, one can see, following the logic of \cite{balkovsky_universal_1999} (see also references therein),  that in the long-time regime, where the inter-particle separation, $s(t)$ is significantly larger than the so-called diffusive scale, $s_d\doteq \sqrt{\kappa\,/\,|\lambda_1|}$ and $\lambda_1(t)\doteq \max_i (\lambda_i\,|\,i=1,\cdots,d)$ is the largest (finite-time) Lyapunov exponent of the Batchelor flow, the fluctuations of ${\bm s}(t)$ are mainly due to the Lyapunov exponents, distributed according to Eq.~(\ref{eq:Cramer}). In other words, in this asymptotic we can approximate the inter-particle distance by
\begin{gather}\label{eq:swim-asymptotic}
   s(t)\approx \exp\Big[\big(\lambda_1(t;0)-\phi\big)t\Big] s_d\,. 
\end{gather}
Substituting $\lambda_1(t;0)$, expressed via $s(t)$ according to Eq.~(\ref{eq:swim-asymptotic}), into Eq.~(\ref{eq:Cramer-1}), and expanding $S_1(\cdot)$ in the Taylor Series around $\bar{\lambda}_1$, we arrive at the following asymptotic expression for statistics of $s(t)=|{\bm s}(t)|$:
\begin{align*}
    \!\!\!& P(s\,|\,t)
    \propto \frac{1}{s}\exp\Big[-t S_1\big(t^{-1}\log\big(s\big/s_d\big)+\phi\big)\Big] \to \frac{1}{s} 
    \\ 
    &\!\!\! \!\times\exp\Big[-t\Big(\!S_1(\bar{\lambda}_1)+\Big(\frac{1}{t}\log\big(s\big/s_d\big)+\phi-\bar{\lambda}_1\Big)^2 S''_1(\bar{\lambda}_1)\Big)\Big]\,. 
\end{align*}
Taking the limit $t\to\infty$ and $s\gg s_d$, leads to
\begin{equation}
P_{\rm st}(s) \propto  \frac{1}{s}\left(\frac{s_d}{s}\right)^{2\,(\phi-\bar{\lambda}_1)\, S''_1(\bar{\lambda}_1)}\,,\label{eq:s_dist}
\end{equation}
where $s_d$, $\bar{\lambda}_1$ and  $S''_1$ are dependent of the system's dynamics. 

Notice, that the stationary version, given in Eq.~\eqref{eq:s_dist}, settles if $\phi>\bar{\lambda}_1$, and that it is fully consistent with Eq.~(\ref{eq:P-sw}), derived for the short $\delta$-correlated velocity gradient, then $S''_1(\bar{\lambda}_1)=1/((d-1)D)$ and $\bar{\lambda}_1{ =d (d-1)D/2}$.

\subsection{BK Model under Proportional Control}\label{sec:steady-Batchelor-Kraichnan} 

The Batchelor-Kraichnan (BK) model is a special case of the Batchelor model where the velocity gradient is not only large scale but is also assumed to be Gaussian and delta-correlated in time (see \cite{falkovich_particles_2001} for the general discussion of the model's historical roots and validity. The BK model in  $d$-dimensions is described by the following pair-correlation function of the velocity gradient matrix, ${\bm \sigma}$, entering Eq.~(\ref{eq:swim-r}):
\begin{align*}
&\mathbb{E}\big[\sigma_{ij}(t)\sigma_{kl}(t')\big]=  D \,(d+1)\,\delta(t-t')\\
   &\times\left(\delta_{jl}\delta_{ik}-\frac{\delta_{ij}\delta_{kl}+\delta_{jk}\delta_{il}}{d+1}\right), \forall i,j,k,l=1,\cdots,d\,, 
\end{align*}
where $\delta(\cdot)$ and $\delta_{ij}$ are the $\delta$-function and the Kronecker symbol respectively. 

Assuming that $\phi$ is sufficiently large we derive from the stochastic ODE~(\ref{eq:swim-r}) the following Fokker-Planck (FP) equation for the spherically symmetric probability density of ${\bm s}$:
(see \cite{chertkov_how_1998} for details):
\begin{align}\label{eq:KFP-swimmers}
&{\cal L}P({\bm s}\,|\,\phi)=0,\\
&{\cal L}=s^{1-d}\frac{\rm d}{{\rm d}s} s^d\left(\phi+\frac{1}{2}\left(D (d-1)s+\frac{\kappa}{s}\right)\frac{\rm d}{{\rm d}s}\right),\nonumber
\end{align}
where $s=|{\bm s}|$; and $\kappa$ stands for variance of the thermal noise in Eq.~(\ref{eq:swim-r}). Solution of Eq.~(\ref{eq:KFP-swimmers}) is 
\begin{equation}\label{eq:P-sw}
P(\bm  s\,|\,\phi) =N^{-1}\bigg(1+\frac{(d-1)Dr^2}{\kappa}\bigg)^{-\phi/(D(d-1))}\,,\nonumber 
\end{equation}
with
\begin{equation*} 
N \equiv \bigg(\frac{\pi\kappa}{(d-1)D}\bigg)^{d/2} \frac{\Gamma\big(\phi/(D(d-1))-d/2\big)}{d\Gamma\big(\phi/(D(d-1))\big)},
\end{equation*}
where $N$ is the normalization coefficient which guarantees that, $\int_0^\infty \Omega_s {\rm d}s P({\bm s}\,|\,\phi)=1$, $\Omega_s=(\pi^{d/2}/\Gamma(d/2+1)) s^{d-1}$.
The solution is valid, i.e. the normalization integral is bounded, if $\phi> (d-1)\, d D/2$.

\subsection{Optimal Stationary Control of BK flow}\label{sec:opt-stat-BK}

The control strategy discussed so far was not optimal, but prescribed. In the case of the BK flow, and as discussed in more details in \cite{chertkov_universality_2023}, we can also find the optimal stationary control --- that is one which solves Eq.~(\ref{eq:SOC}) at $T\to\infty$ in the case of no penalty for the special case of the (negative) reward function
\begin{gather}\label{eq:reward}
r\left({\bm s}(t'),{\bm a}(t')\right):=-\big\|{\bm a}(t')\big\|_2^{2} -\beta\, \big\|{\bm s}(t')\big\|_2^{2},
\end{gather}
most suitable for the particle problem -- where we balance of the quadratic cost of action with the quadratic cost penalizing quadratically for increase in the inter-particle distance. 

Considering the steady state (infinite horizon) and no discount and utilizing Eq.~(\ref{eq:P-sw}) while also assuming that $\phi>\phi^{(s)}=\tilde{D}/2$, where $\tilde{D}=D (d+2)(d-1)$, we arrive at 
\begin{align}\label{eq:SOC-steady}
\phi^*  &=\text{arg}\min_{\phi} \bigg[(\phi^2+\beta) \int_0^\infty s^2\, P(s|\phi)\, {\rm d}s\bigg]\nonumber\\
&= \text{arg}\min_{\phi} \frac{\phi^2+\beta}{2\phi-\tilde{D}}= \frac{\tilde{D}\!+\!\sqrt{4\beta\!+\!\tilde{D}^{\,2}}}{2}\,.
\end{align}

\section{State Value function in the Batchelo-Kraichnan Flow}\label{sec:value-SK}

In this section, we use the BK theory to evaluate the state value function. Significance of this result is in its further utilization in the next section as an estimate for the physicist (replacing critic) bias in the PC RL algorithm.  

Let's consider the continuous time version of Eq.~(\ref{eq:V}); the value function which we evaluate over a finite horizon from $t$ to $T$, with the discount factor $\nu$, reads
\begin{align}\label{eq:bias}
&V_\phi\big(t,{\bm s}(t)\big) = (\beta + \phi^2)\int_t^T {\rm d}t' e^{-\nu (t'-t)}\bigg(e^{-2\phi(t'-t)} \nonumber\\
&\times\mathbb{E}\Big[\big\|\bm W(t';t)\, \bm s\big\|_2^2\Big]
+\kappa\int_{t}^{t'} {\rm d}t'' e^{-2\phi(t'-t'')} \nonumber\\
& \times\mathbb{E}\Big[\text{Tr}\big[{\bm W}^{\top}(t';t'')\, {\bm W}(t';t'')\big]\Big]\bigg),
\end{align}
where ${\bm W}$ satisfies Eq.~(\ref{eq:W}); the expectations are over the statistics of ${\bm \sigma}$, which is not yet specified, and over the white-Gaussian thermal forces. 
Here, in Eq.~(\ref{eq:bias}), we evaluate the state value function conditioned to a value of the control coefficient $\phi$ fixed to a constant, and since $\phi$ is the only parameter in PC, we will label $V$ with it.

In the case of the SK flow we observe that $\mathbb{E}\big[W_{ki}(t';t)W_{kj}(t';t)\big]=\delta_{ij}\times\exp\big(-(t'-t)\,\tilde{D}\big)$. This results in the following  closed-form expression for Eq. \eqref{eq:bias}
\begin{align} \label{eq:b-BK-1}
&V_\phi\big(t,s(t)\big) = (\beta+\phi^2)\int_t^T {\rm d}t' e^{-\nu (t'-t)}\bigg(s^2 e^{-(2\phi-\tilde{D})(t'-t)} \nonumber \\
&+\kappa d\int_{t}^{t'} {\rm d}t'' e^{-(2\phi-\tilde{D})(t'-t'')}\bigg) =: B(t)\, s^2 +C(t),
\end{align}
with
\begin{align*}
B(t)  &= \frac{(\beta+\phi^2)\big(1-e^{-(T-t)(\nu+2\phi-\tilde{D})}\big)}{\nu+2\phi-\tilde{D}},\nonumber \\ 
C(t) &= \frac{d\kappa (\beta+\phi^2)}{\nu(2\phi-\tilde{D})}\big( 1-e^{-\nu(T-t)}\big)- \frac{d\kappa}{(2\phi-\tilde{D})}\, B(t) .\nonumber 
\end{align*}
It is this expression which we will use as a baseline for the ABC flow in Section \ref{sec:results}, after estimating respective value of $\tilde{D}$ in Section \ref{sec:ABC}, hence providing physics-based guidance to the $critic$ block of the method, as shown in Fig.~\ref{fig:ac_cartoon}.

Some remarks are in order.  First, notice that in the infinite horizon, $T\to\infty$, and then no discount, $\nu\to\infty$, limits the surviving, $B$, part of the final expression in Eq.~(\ref{eq:b-BK-1}), is fully consistent with yet not optimized over $\phi$ part of the steady expression in Eq.~(\ref{eq:SOC-steady}). Second, note that even though we emphasize in Eq.~(\ref{eq:b-BK-1}) the dependence of $A$ and $B$ only on $t$, the formula actually provides explicit dependence on all other parameters of the BK flow and of the optimization formulation. Third, the expression in Eq.~(\ref{eq:b-BK-1}) for $V_\phi$ applies only to cases where the statistics of separation stabilize over time, requiring that $2\phi > \tilde{D}$. Moreover, we expect a critical slowdown effect -- when approaching the boundary, i.e., $\phi=\tilde{D}/2+\varepsilon$ and $\varepsilon \to 0$, the time required to establish the steady distribution of the separation diverges. This means that, given practical limitations on collecting statistics, Eq.~(\ref{eq:b-BK-1}) does not apply in the regime of small positive $\varepsilon$, potentially leading to statistics of the separation which does not stabilize at $T\to \infty$. In this case we choose to be conservative and work with $\phi$ which are larger than $\tilde{D}/2$ with a reserve.

\section{Swimming in Arnold–Beltrami–Childress Flow}\label{sec:ABC}

To validate our RL approaches we use the Arnold–Beltrami–Childress (ABC) flow~\cite{zhao_chaotic_1993} in the chaotic regime, where trajectories diverge dramatically, largely in an exponential fashion. However, in this case, the assumptions of the BK model are not exactly satisfied and the equation \eqref{eq:b-BK-1} do not provide an exact match and can only serve as a rough approximation. 

\begin{figure}[h!]
    \includegraphics[width=\linewidth]{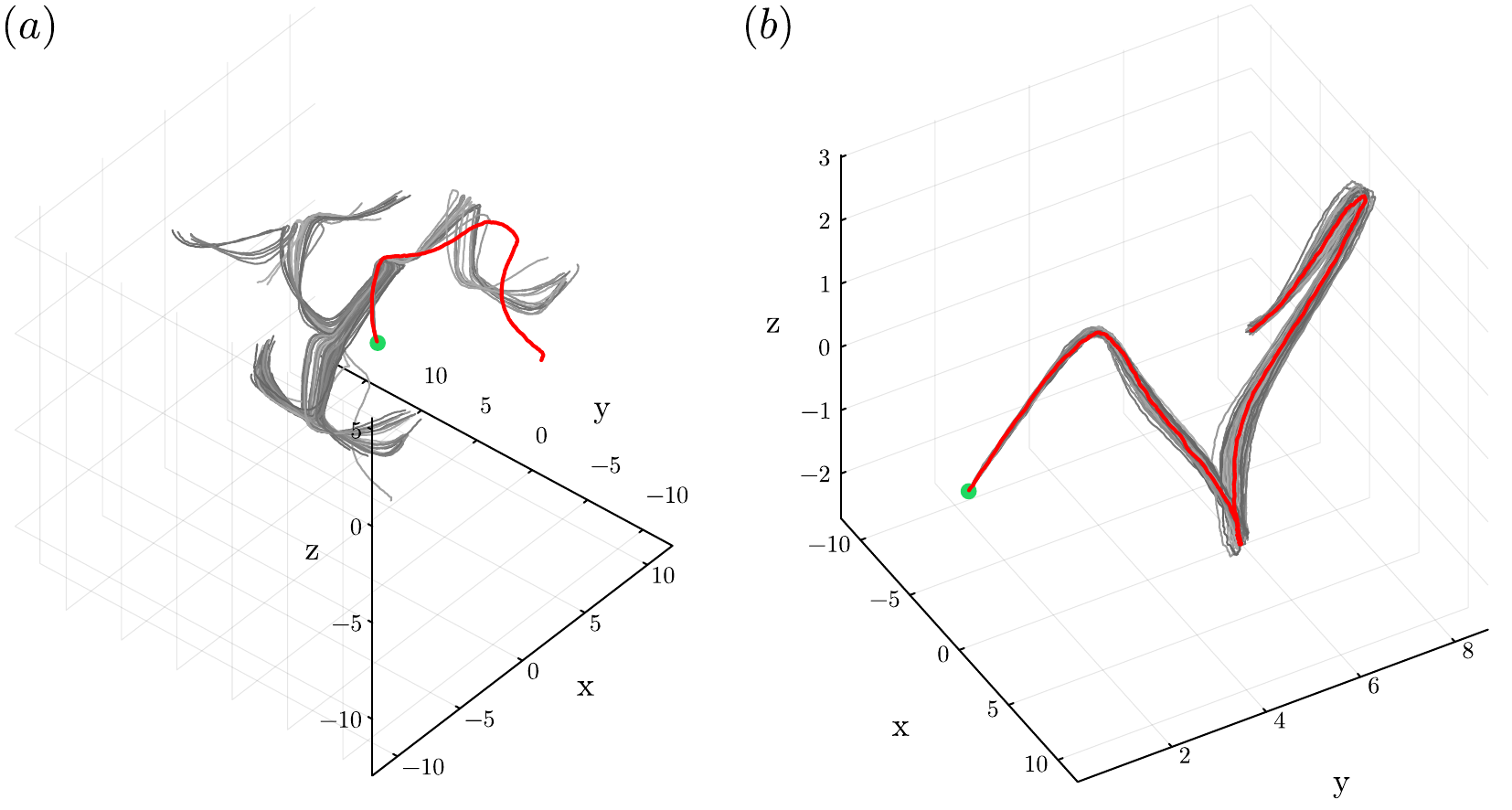} 
    \caption{Sample trajectories of an active particle (gray) and its passive target (red) in an ABC flow. (a) No control: trajectories starting from the same point diverge chaotically. (b) The case of PC control: the active particle closely follows the trajectory of the target particle. While trajectories are not identical, their divergences from the passive target is far less than in the uncontrolled case.
    }
    \label{fig:abc_trajectories}
\end{figure}

This flow is a three-dimensional incompressible velocity field which is an exact solution of Euler's equation. Its representation in Cartesian coordinates, augmented with an weak additive noise is  
\begin{align*}
v_{\alpha,x}({\bm s}_{\alpha}) &=A \sin\big(s_{\alpha,z}\big)+C \cos\big(s_{\alpha,y}\big)+\xi_{\alpha,x},
\\
v_{\alpha,y}({\bm s}_{\alpha}) &=B \sin\big(s_{\alpha,x}\big)+A \cos(s_{\alpha,z})+\xi_{\alpha,y},\\ 
v_{\alpha,z}({\bm s}_{\alpha}) & =C \sin\big(s_{\alpha,y}\big)+B \cos\big(s_{\alpha,x}\big)+\xi_{\alpha,z}, 
\end{align*}
with 
\begin{align*}
\xi_{\alpha,x}=\sqrt{\kappa}\, \frac{{\rm d}w_x}{{\rm d}t},\quad
\xi_{\alpha,y}=\sqrt{\kappa}\, \frac{{\rm d}w_y}{{\rm d}t},\quad
\xi_{\alpha,z}=\sqrt{\kappa}\, \frac{{\rm d}w_z}{{\rm d}t},
\end{align*}
where $\alpha=1,2$ is the particle index (for the active particle and its passive counterpart, resepctively) and $A,B$ and $C$ are three flow parameters and $w_x,w_y,w_z$ are unit variance Wiener processes.  With appropriate selection of flow parameters the flows are chaotic so that if a pair of passive particles are placed in the flow their separation grows exponentially with time in the $s\ll 2\pi$ regime, as illustrated in Fig.~(\ref{fig:abc_trajectories}). The small additive noise strength $\kappa$ is introduced to provide a change for divergence even if the active particle's and its target's initial positions coincide.

\begin{figure}
\center
\includegraphics[width=0.85\columnwidth]{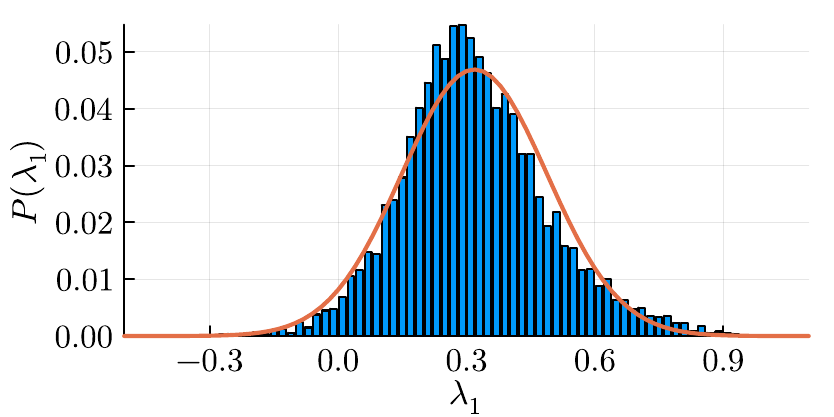}
\caption{Finite time statistics of the leading Lyapunov exponent $\lambda_1(t)\approx\log(\bm W^\top(t;0)\bm W(t;0))/t$ in the ABC flow, evaluated at $t$ equal to the episode's duration.  
\label{fig:ABC}}
\end{figure}

Probability distribution of sampled finite time leading Laypunov exponent, $\lambda_1(t)\approx\log(\bm W^\top(t;0)\bm W(t;0))/t$, in the ABC flow for $A=1.,\ B=0.7,\ C=0.43$ and evaluated at $t$ equal to the duration of the episode is shown in Fig.~(\ref{fig:ABC}). In this case relation between the $\tilde{D}$ parameter, measuring the eddy-diffusivity strength of the BK flow, and the average leading Lyapunov exponent is $\tilde{D} = 2\Bar{\lambda}_1(1+2/d)$. This relation was subsequently utilized in Eq.~(\ref{eq:b-BK-1}).

\section{Numerical Experiments} \label{sec:results}

We first show, in Section \ref{sec:val}, that an agent following a proportional control do indeed reproduce what the theory predicts. We then test the AP agent and demonstrate its superiority over the state-of-the-art AC agents in Section \ref{sec:comp_w_ac}, as well as compare it to a prescribed control with a near-optimal swimming rate in Section \ref{sec:comp_w_prescribed}. 

When comparing the RL approaches -- the AP agent versus the AC agents -- we use the reward defined by Eq.~(\ref{eq:reward}) with a weighting factor of $\beta = 0.1$. This choice was tuned to the flow conditions, allowing both components of the reward -- maintaining proximity to the target (passive particle) and minimizing control efforts (energy expenditure) -- to be clearly expressed and optimized simultaneously, resulting in a pronounced and effective trade-off.

\subsection{Validation of the Physicist (Baseline)}\label{sec:val}

\begin{figure}[ht]
    \centering
        \includegraphics[width=\columnwidth]{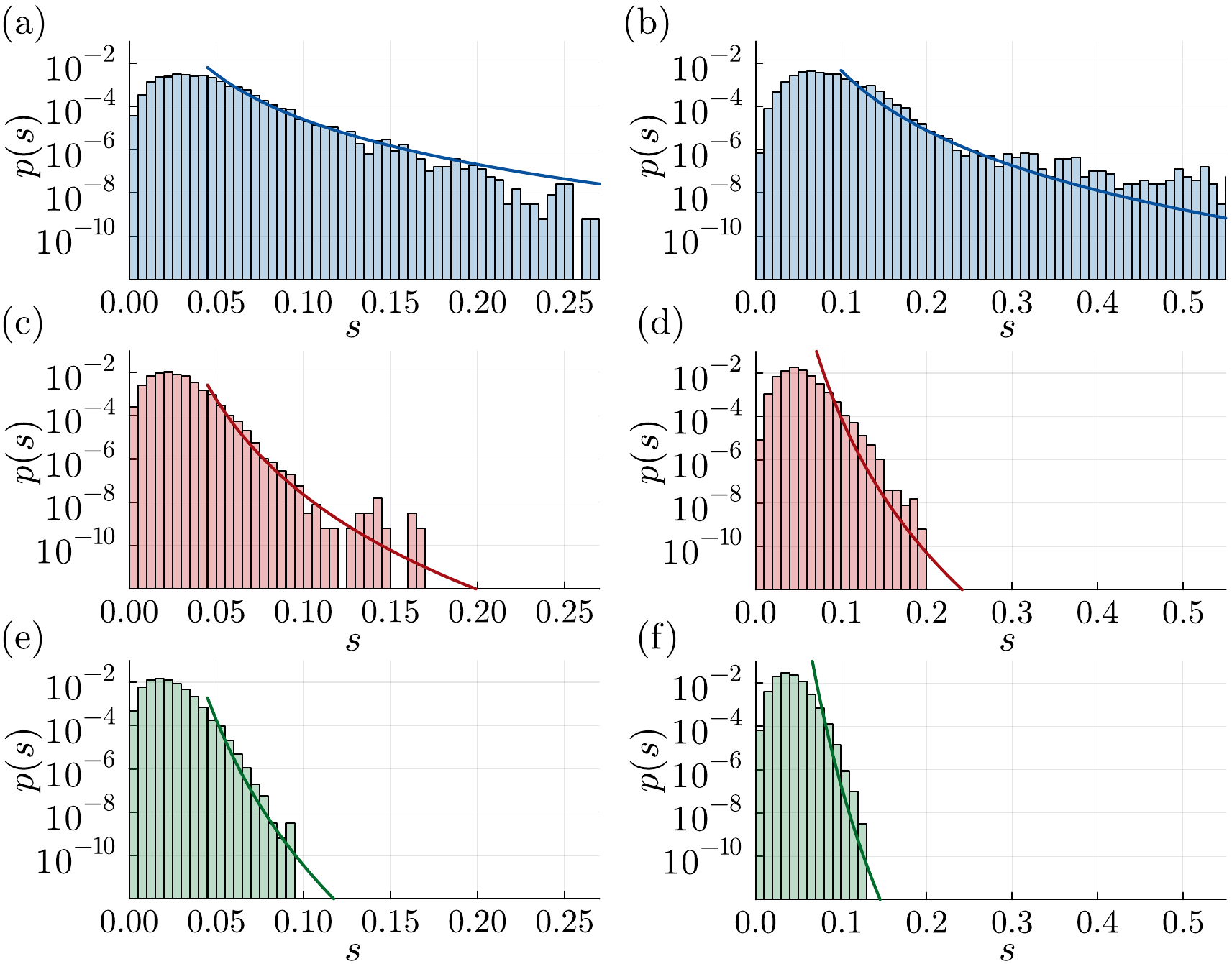}
    \caption{Distribution of separation $\bm{s}(t)$ for prescribed control with fixed values of $\phi$: $0.6$ (blue), $1.1$ (red), and $1.6$ (green) for B-K flows (left column) and ABC flows (right column). The lines represent the predictions of Eq.~\eqref{eq:s_dist}. For the ABC flow, $\bar{\lambda}_1$ was obtained from simulations, as shown in Fig.~\ref{fig:ABC}.
    }
    \label{fig:s_dist}
\end{figure}

Fig.~\ref{fig:s_dist} compares the theoretical predictions from Eq.~\eqref{eq:s_dist} with simulation results. We observe good qualitative agreement between theory and simulations for both the BK and ABC flows. The agreement is slightly better in the BK case, where both the SOC formulation and the value function $V_\phi$ are analytically tractable, compared to the ABC flow, where the estimation of $\bar{\lambda}_1$ is approximate.

\begin{figure}[ht]
    \centering
    \includegraphics[width=\columnwidth]{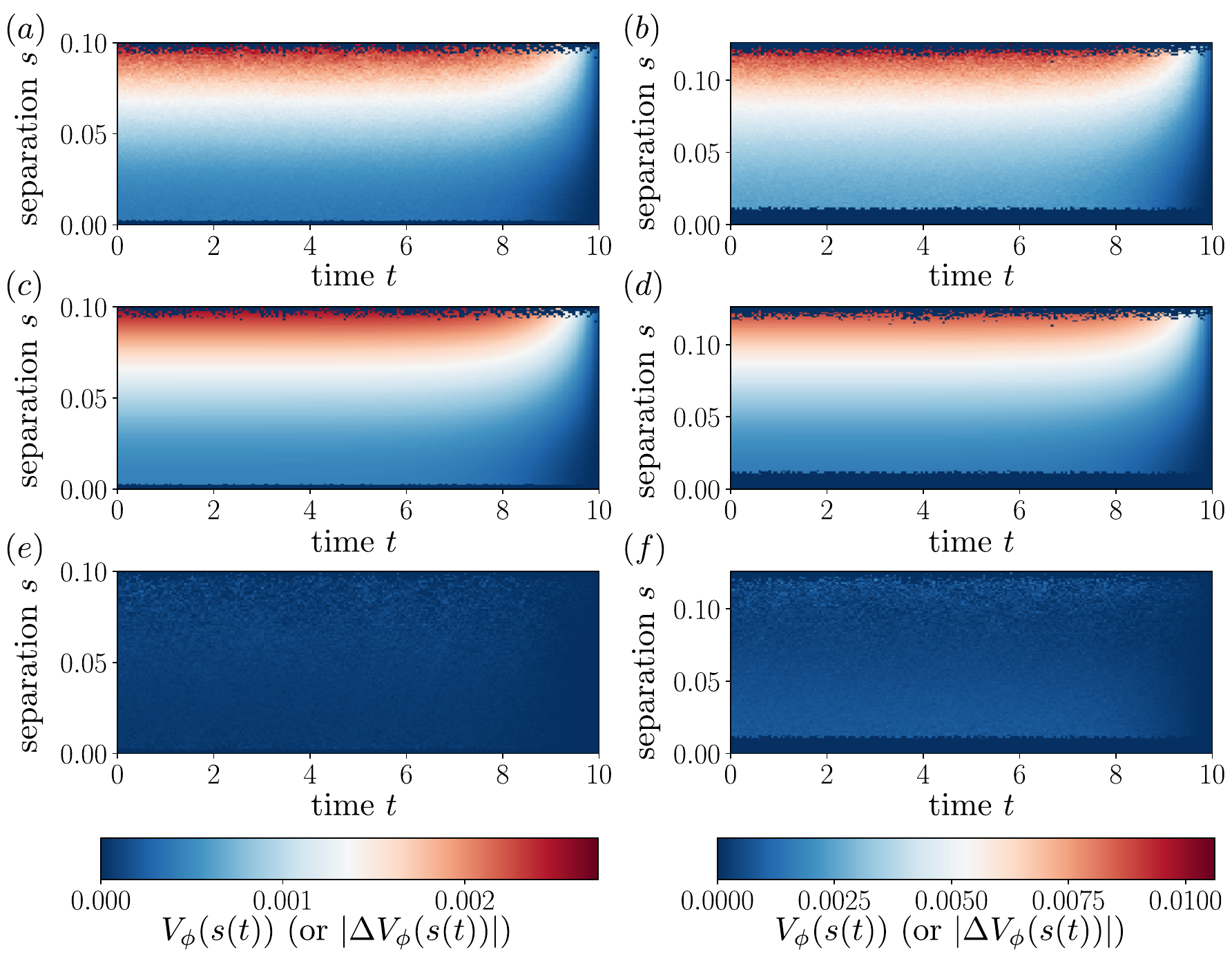} 
    \caption{Panels (a)-(b) display the expected return, $V_\phi(s(t))$, observed in simulations. The corresponding theoretical values, as predicted by Eq.~\eqref{eq:b-BK-1}, are shown in panels (c)-(d), and the difference between simulation and theory is presented in panels (e)-(f). The left column -- panels (a), (c), and (e) -- corresponds to the BK flow with $\phi=0.574$, while the right column -- panels (b), (d), and (f) -- corresponds to the ABC flow with $\phi=1.1$. (See the main text for further details on the choice of $\phi$ values in this validation study.)  }\label{fig:return_val}
\end{figure}

The agreement shown in Fig.~\ref{fig:s_dist} naturally leads us to the next step -- a more detailed comparison of theory versus simulations for the value function, as presented in Fig.~\ref{fig:return_val}. In that figure, we compare the empirical and theoretical values of the value function corresponding to the discounted return in Eq.~\eqref{eq:SOC}, with the predicted baseline in Eq.~\eqref{eq:b-BK-1} or, more precisely, with the (properly normalized) expected return $G$, defined in Eq.~\eqref{eq:avg_return}. The predictions for the ABC flow match surprisingly well, with discrepancies between simulation and theoretical values only slightly larger than in the BK model. In Fig.~\ref{fig:return_val}, we present results for a single value of $\phi$, chosen to be near its asymptotically optimal value, $\phi_*$; however, similar agreements were observed for other values of $\phi$ (not shown), as anticipated from the agreement seen in Fig.~\ref{fig:s_dist}.

\subsection{Comparison with Standard Actor-Critic Methods}\label{sec:comp_w_ac}

Now that our physics baseline, described by Eq.~\eqref{eq:b-BK-1}, has been validated, we proceed in this section to embed it into the critic, while keeping the actor tuned as before --using a NN. We refer to this combination of a physics-based critic and an NN-based actor as the Actor-Physicist (AP) agent/algorithm. Our focus is on comparing the performance of the AP agent with that of standard AC agents, specifically those using A2C \cite{mnih_asynchronous_2016} and PPO \cite{schulman_proximal_2017} schemes.

\begin{figure}[ht]
    \centering
        \includegraphics[width=0.8\columnwidth]{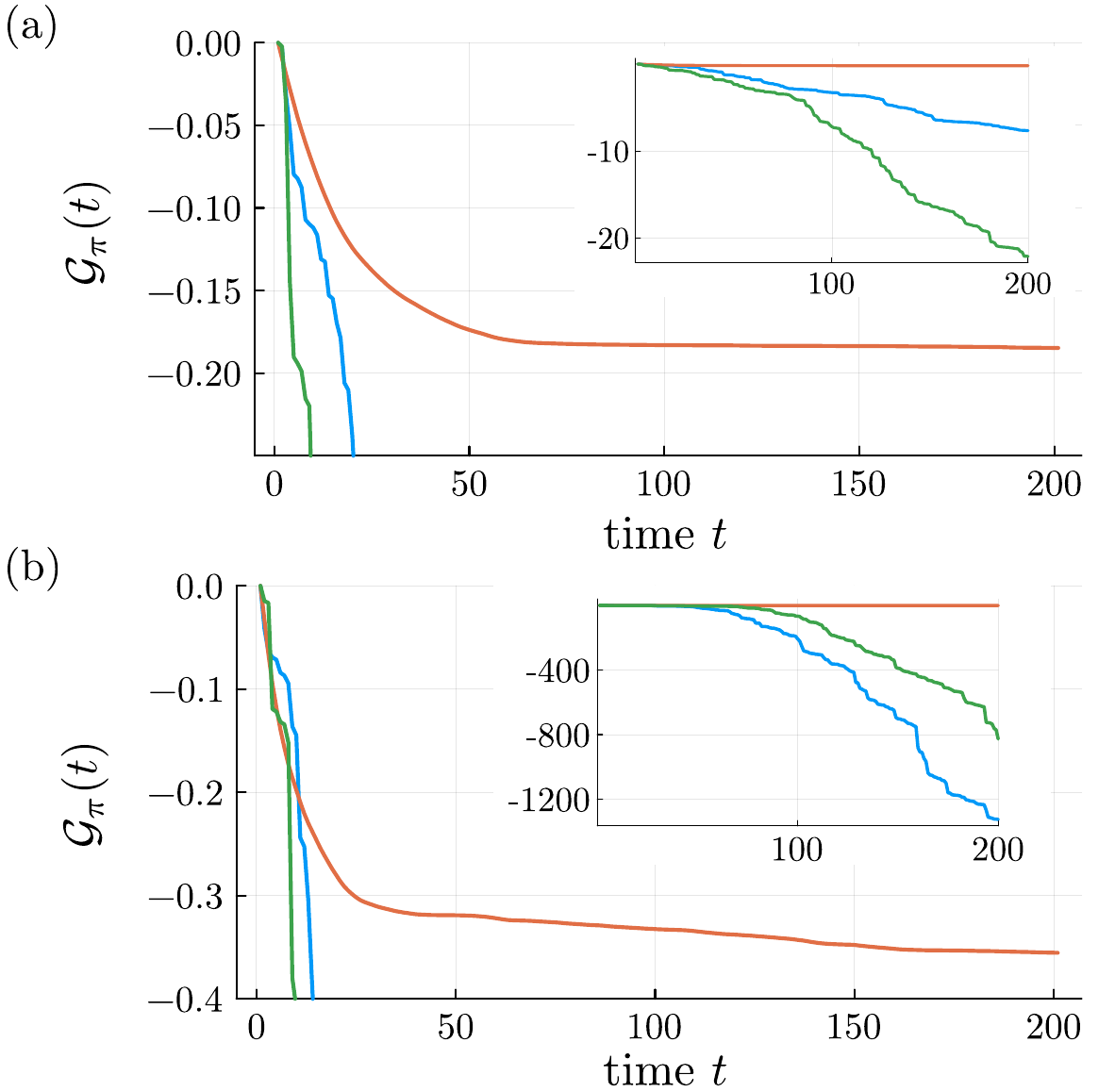}
    \caption{ Post-training average return $\mathcal{G_\pi}$, defined in Eq.\eqref{eq:avg_return}, for AP algorithm (red) and  AC algorithm with the A2C scheme (blue) and with the PPO scheme (green). The averaging is over 500 samples. The results shown correspond to the optimal $\phi^*=0.574$ in the case of the BK flow (a) and estimate of the optimal $\phi^*=1.1$ in the case of the ABC flow (b). 
    }
       \label{fig:sep}
\end{figure}

Fig.~\ref{fig:sep} (a) and (b) demonstrate that the AP agent consistently outperforms the AC agents across all our experiments, with particularly notable results in the ABC flow scenarios. Specifically, the AC agents using both A2C and PPO did not yield meaningful outcomes within 250 training trials, as they failed to converge, whereas the AP agent achieved reliable convergence within the same number of trials. We attribute the poor performance of standard AC agents to the fact that, while theoretically optimal in the asymptotic limit, practical implementations are inherently non-asymptotic, especially in terms of the limited number of trials available for training. Additionally, we hypothesize that the failure of A2C and PPO may be due to the non-standard, fat-tailed statistics of our problem, characterized by extended algebraic tails and significant variability in reward values across different flow configurations, spanning several orders of magnitude. Such conditions are known to present challenges for standard AC methods \cite{acrossvalues}.

We want to emphasis that, we do not conclude that AC agents cannot learn to swim in a turbulent flow at all, but that, with limited observation and training resources, they fail, whereas our AP agent is able to learn how to perform this task under the same conditions.

\subsection{Comparison with Prescribed Control}\label{sec:comp_w_prescribed}

Now that we have shown that the AP agent outperforms standard AC agents. We would like to check if our AP agent actually performs better than the prescribed control it is built upon. To investigate this we compare these two control schemes in different regimes.

Table~\ref{tab:comp_w_pc} shows a comparison between our AP agent and PC for different values of $\phi$. We observe that the trained AP agents tend outperform the corresponding  PC strategies. For both flows, there is one exception which is when the value of the control is set (close) to the asymptotically derived optimal $\phi^*$. 

\setlength{\tabcolsep}{0pt} 
\setlength{\arrayrulewidth}{1pt} %
\begin{table}[ht]
\begin{tabular}{lll}
\hline
(a)\vphantom{$\intop_p^f$}&\multicolumn{2}{c}{$\mathbb{E}_{s_0\sim\mathcal{D}}\big[V(s_0)\big]$}\\
\hline
$\phi$ \hphantom{asdf}  & PC  $\phi s(t)$ \hphantom{asdf} & AP \\
0.3 & -0.21477 & -0.18348\\
\rowcolor{red!40}0.574 & -0.17589 & -0.18143\\
0.9 & -0.21680 & -0.18953\\
1.2 & -0.26607 & -0.17662\\
\hline
\end{tabular}\hspace{6pt}
\begin{tabular}{lll}
\hline
(b)\vphantom{$\intop_p^f$}&\multicolumn{2}{c}{$\mathbb{E}_{s_0\sim\mathcal{D}}\big[V(s_0)\big]$}\\
\hline
$\phi$\hphantom{asdf} & PC  $\phi s(t)$\hphantom{asdf} & AP \\
0.6 & -0.40784 & -0.34884\\
\rowcolor{red!40} 1.1 & -0.32381 & -0.36321\\
1.6 & -0.40041 & -0.31272\\
2.1 & -0.49141 & -0.34564\\
\hline
\end{tabular}
\caption{
This figure compares the Prescribed Control (PC) case with the AP agent/algorithm at varying $\phi$ values for the BK flow (subfigure (a)) and the ABC flow (subfigure (b)). The results presented are averages over 1,000 randomly drawn initial states $\bm{s}_0$. The highlighted rows indicate the optimal $\phi^*$ for (a) and the estimated optimal $\phi^*$ for (b) used in the simulations.} 
\label{tab:comp_w_pc}
\end{table}

\begin{figure}[ht]
    \centering
        \includegraphics[width=0.8\columnwidth]{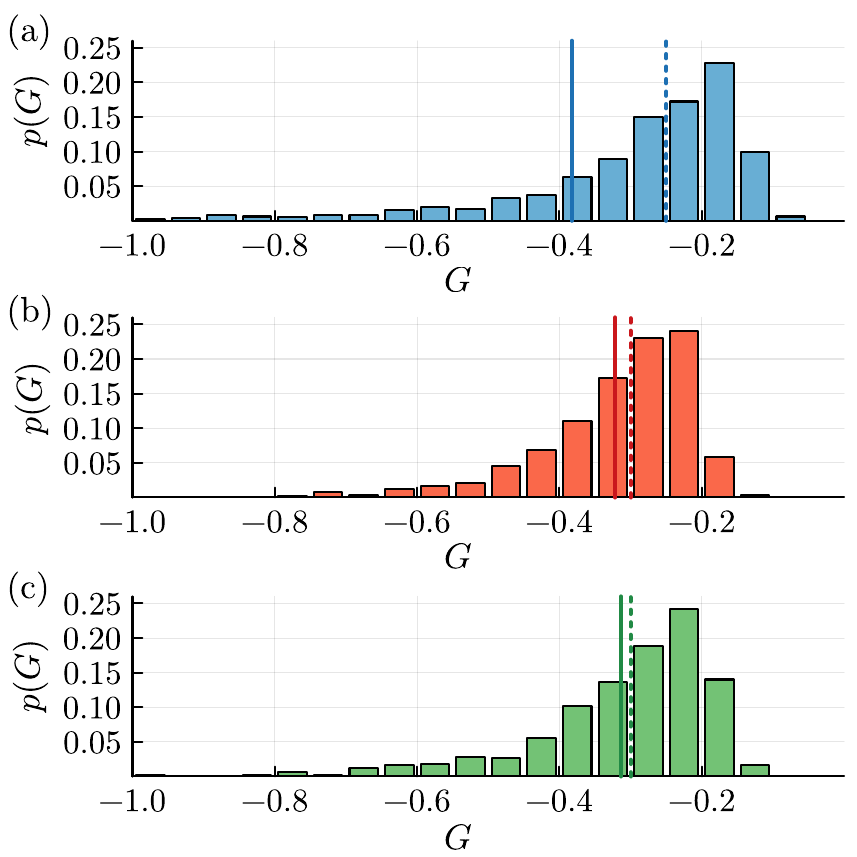}
    \caption{Distributions of the return, $G:=\sum_{k=0}^N\gamma^kr(s_k,a_k)$, in an ABC flow  with $\phi=1.1$ for the AP agent  (a), prescribed control (b) and the hybrid scheme (c). The mean and median are displayed as solid and dashed lines.}
    \label{fig:return}
\end{figure}

Results presented in Table~\ref{tab:comp_w_pc} demonstrates that if the optimal value $\phi^*$ of the control can be determined accurately, the prescribed control with the optimal value performs better than our AP algorithm. However, the examination of the distributions of returns, shown in Fig.~\ref{fig:return} (a) and (b), suggests that the lower average return of the AP is due to large, but infrequent failures. Theses panels also show that the median performance of the AP is higher than the median performance of the fixed $\phi$ agent. This means that the trained AP is most likely to outperform its baseline on a particular realization. The higher rate of extreme cases under the control of the AP incurs very large penalties in particular realizations.

\begin{figure}[ht]
    \centering
        \includegraphics[width=0.8\columnwidth]{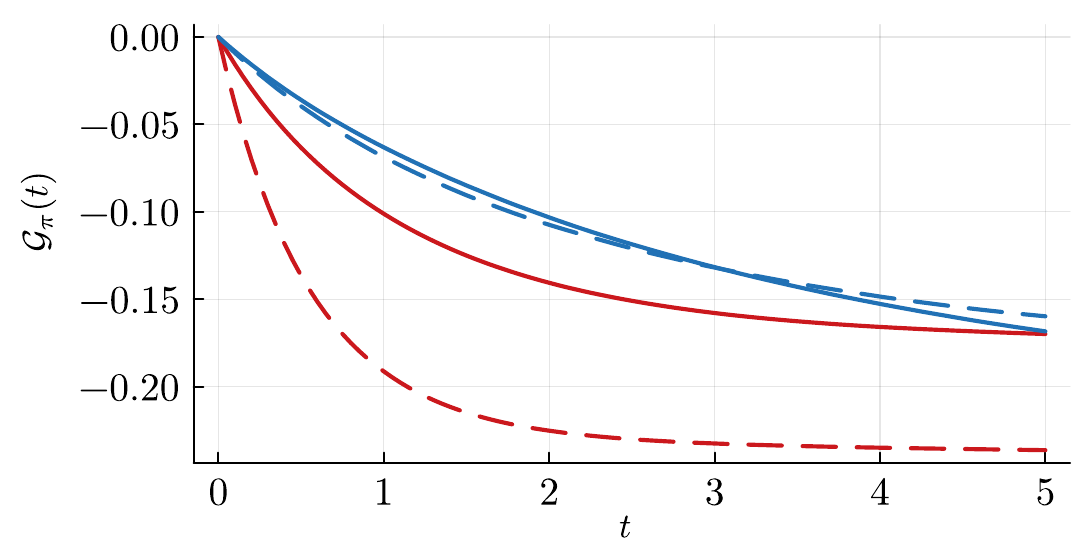}
    \caption{Average return $\mathcal{G_\pi}$ -- defined in Eq.~\eqref{eq:avg_return} and computed through averaging over 250 realizations -- shown for the trained AP agent (blue) and prescribed control (red) with asymptotically optimal $\phi^*$ (solid) and suboptimal $\phi$ (dashed).  }  \label{fig:short}
\end{figure}

An interesting result is reported in Fig.~\ref{fig:short} when we focus on the performance of the trained agent over shorter time horizons. In this case, the trained agent tends to outperform its baseline even with $\phi^*$ selected according to Eq.~(\ref{eq:SOC-steady}). This makes sense as the optimal strength $\phi^*$ was optimized for an asymptotic stationary distribution of a particle under a proportional control. If the initial separation is distributed such that it is often  improbable (as evaluated with the stationary, that is idealized, distribution), PC with $\phi^*$  is not the best possible action and the AP agent is able to prevail in this context.

The mention of typical versus atypical behavior in the challenging case where the AP agent does not perform as well as the fixed $\phi$ agent on average helps us draw useful lessons. We learn that a major benefit of a physics-informed baseline is the ability to monitor the trained agent's performance and identify when the agent may be falling into a regime it cannot handle. In such regimes, intervention with a naive fixed $\phi$ strategy -- that is, designing a hybrid policy that mixes or explores in parallel the RL and fixed $\phi$ strategies -- can avoid catastrophic failures and return the agent to a regime where it outperforms the fixed $\phi$ policy. Even a simple hybrid solution that switches to the fixed $\phi$ strategy if the average advantage of the previous $n$ training steps is below a threshold shows improvements. Our results, illustrated in Fig.~\ref{fig:return}, show improvements even with arbitrary choices for $n$ and the threshold, suggesting that significant further enhancements of the approach can be achieved through more principled optimization.

\begin{figure}[ht]
    \centering
        \includegraphics[width=\columnwidth]{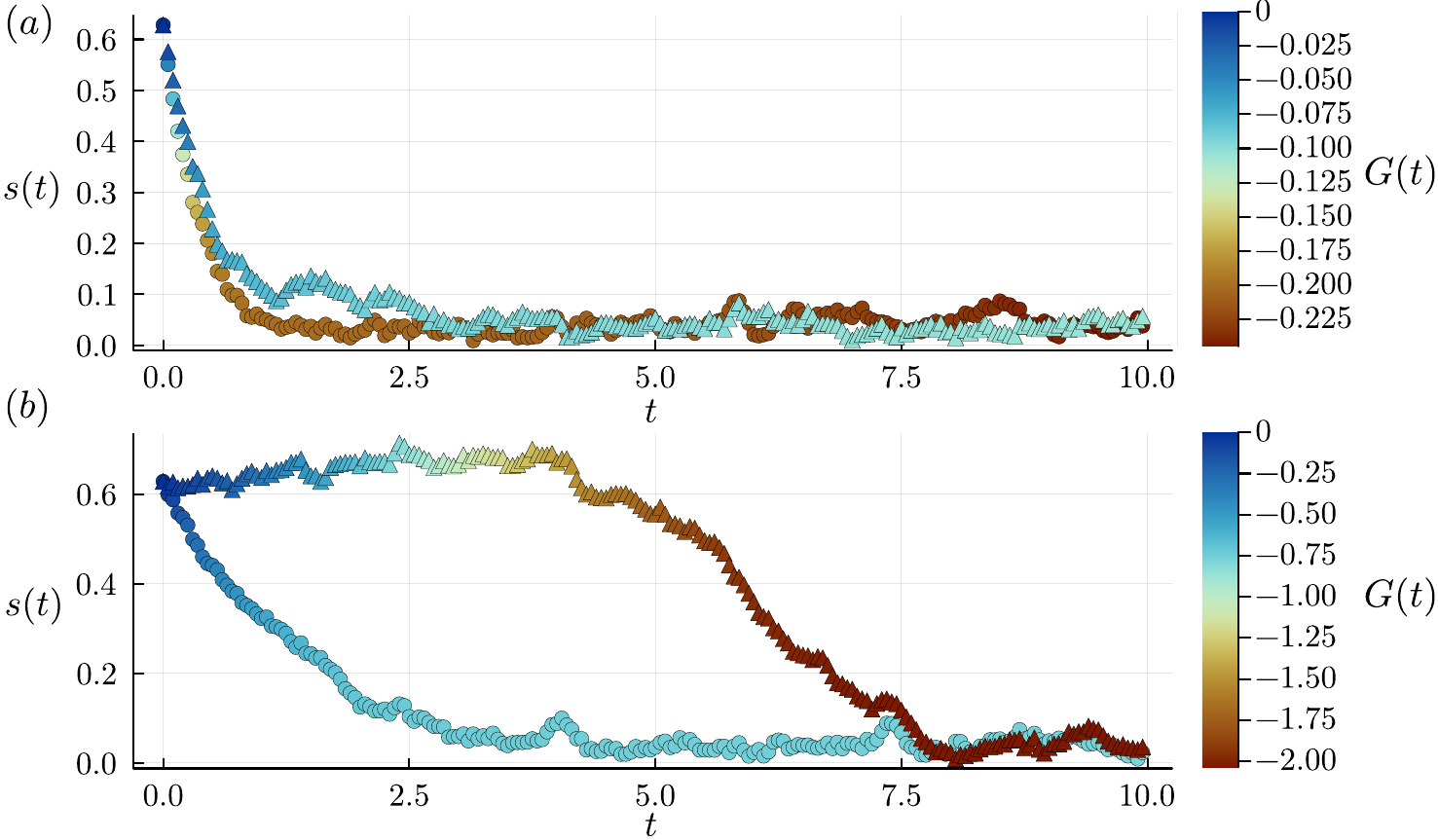}
    \caption{Selected separation of trajectories for AP (triangle) and PC (circle): typical (a) and atypical (b), the former meaning from the bulk of the distribution in Fig.~\ref{fig:return} and the latter meaning from its tail.  Makers' colors display the return, $G(t):=\sum_{k=0}^n\gamma^kr(s_k,a_k)$, with $n=t/\Delta$.}
    \label{fig:pen_n_sep}
\end{figure}

To further elaborate on the nuances of typical versus atypical flows, Fig.~\ref{fig:pen_n_sep} shows sample trajectories of the evolution of the separation and the penalty accrued in a typical trajectory versus an atypical trajectory. In the typical case, the AP agent is much more efficient than the agent following the fixed $\phi$ policy. In the atypical case, the AP agent is overly energy conservative and thus converges to the passive particle slowly, even showing divergence from the passive partner at the start. In this case, the high penalty is accrued due to the cost of separation. These results imply that the agent could greatly benefit from expanding the state space since it currently has no way to distinguish the typical and atypical episodes shown in Fig.~\ref{fig:pen_n_sep}.

\section{Conclusion} \label{sec:discussions}

We study the challenging problem of controlling the separation between a swimmer and its passive target in chaotic flows using reinforcement learning. Initially, we observe that state-of-the-art deep NN-based RL algorithms of the actor-critic (AC) type fail at this task due to their inability to handle fat-tailed statistics, which exhibit extended power-law or algebraic tails. To address this, we propose a physics-informed approach where the critic utilizes analytical estimates, specifically the average maximum Lyapunov exponent of the flow.

We demonstrate the effectiveness of this approach on the BK flow and validate it on the ABC flow. This method replaces NN approximations of value functions with flow-control-informed functions, reducing the computational burden while incorporating domain-specific knowledge and maintaining convergence guarantees. Even imperfect physics-informed functions can improve policy training, provide interpretable insights into policy failures, and serve as an effective alarm mechanism.

While this study focuses on controlling a pair of particles, the approach naturally extends to monitoring and controlling groups navigating turbulent flows, such as air drones, bird flocks, or aquatic drone swarms. Some underlying multi-agent problems may be simplified to a $k$-nearest neighbors plus mean-field control framework, as discussed in \cite{borra_optimal_2021}. Combining the approach developed in this manuscript with the mean-field framework and other notable methods, such as \cite{novati_automating_2021} and \cite{bae_scientific_2022}, presents significant opportunities for advancing the emerging field of multi-agent reinforcement learning. A particularly intriguing direction is the control of swarms of Lagrangian particles with contrasting objectives, such as maintaining group cohesion, avoiding close encounters, and minimizing energy expenditure for maneuvering.

\section*{Acknowledgments}
We acknowledge R. Ferrando for useful discussion. This work was supported by a start up funding from the University of Arizona and a subcontract from Los Alamos National Laboratory.

\section*{Code Availability}
The code that was used for this work and, in particular, this environment is available at \url{https://github.com/Cfckoh/RL_swimmers/tree/main}.

\bibliography{prr_references}

\begin{thebibliography}{32}%
\makeatletter
\providecommand \@ifxundefined [1]{%
 \@ifx{#1\undefined}
}%
\providecommand \@ifnum [1]{%
 \ifnum #1\expandafter \@firstoftwo
 \else \expandafter \@secondoftwo
 \fi
}%
\providecommand \@ifx [1]{%
 \ifx #1\expandafter \@firstoftwo
 \else \expandafter \@secondoftwo
 \fi
}%
\providecommand \natexlab [1]{#1}%
\providecommand \enquote  [1]{``#1''}%
\providecommand \bibnamefont  [1]{#1}%
\providecommand \bibfnamefont [1]{#1}%
\providecommand \citenamefont [1]{#1}%
\providecommand \href@noop [0]{\@secondoftwo}%
\providecommand \href [0]{\begingroup \@sanitize@url \@href}%
\providecommand \@href[1]{\@@startlink{#1}\@@href}%
\providecommand \@@href[1]{\endgroup#1\@@endlink}%
\providecommand \@sanitize@url [0]{\catcode `\\12\catcode `\$12\catcode
  `\&12\catcode `\#12\catcode `\^12\catcode `\_12\catcode `\%12\relax}%
\providecommand \@@startlink[1]{}%
\providecommand \@@endlink[0]{}%
\providecommand \url  [0]{\begingroup\@sanitize@url \@url }%
\providecommand \@url [1]{\endgroup\@href {#1}{\urlprefix }}%
\providecommand \urlprefix  [0]{URL }%
\providecommand \Eprint [0]{\href }%
\providecommand \doibase [0]{https://doi.org/}%
\providecommand \selectlanguage [0]{\@gobble}%
\providecommand \bibinfo  [0]{\@secondoftwo}%
\providecommand \bibfield  [0]{\@secondoftwo}%
\providecommand \translation [1]{[#1]}%
\providecommand \BibitemOpen [0]{}%
\providecommand \bibitemStop [0]{}%
\providecommand \bibitemNoStop [0]{.\EOS\space}%
\providecommand \EOS [0]{\spacefactor3000\relax}%
\providecommand \BibitemShut  [1]{\csname bibitem#1\endcsname}%
\let\auto@bib@innerbib\@empty
\bibitem [{\citenamefont {Chertkov}(2023)}]{chertkov_universality_2023}%
  \BibitemOpen
  \bibfield  {author} {\bibinfo {author} {\bibfnamefont {M.}~\bibnamefont
  {Chertkov}},\ }\href {http://arxiv.org/abs/2303.09635} {\bibinfo {title}
  {Universality and {Control} of {Fat} {Tails}}} (\bibinfo {year} {2023}),\
  \bibinfo {note} {arXiv:2303.09635 [cond-mat, physics:nlin, stat]}\BibitemShut
  {NoStop}%
\bibitem [{\citenamefont {Sutton}\ and\ \citenamefont
  {Barto}(2018)}]{sutton_reinforcement_2018}%
  \BibitemOpen
  \bibfield  {author} {\bibinfo {author} {\bibfnamefont {R.~S.}\ \bibnamefont
  {Sutton}}\ and\ \bibinfo {author} {\bibfnamefont {A.~G.}\ \bibnamefont
  {Barto}},\ }\href@noop {} {\emph {\bibinfo {title} {Reinforcement learning:
  an introduction}}},\ \bibinfo {edition} {second edition}\ ed.,\ Adaptive
  computation and machine learning series\ (\bibinfo  {publisher} {The MIT
  Press},\ \bibinfo {address} {Cambridge, Massachusetts},\ \bibinfo {year}
  {2018})\BibitemShut {NoStop}%
\bibitem [{\citenamefont {Mnih}\ \emph {et~al.}(2016)\citenamefont {Mnih},
  \citenamefont {Badia}, \citenamefont {Mirza}, \citenamefont {Graves},
  \citenamefont {Lillicrap}, \citenamefont {Harley}, \citenamefont {Silver},\
  and\ \citenamefont {Kavukcuoglu}}]{mnih_asynchronous_2016}%
  \BibitemOpen
  \bibfield  {author} {\bibinfo {author} {\bibfnamefont {V.}~\bibnamefont
  {Mnih}}, \bibinfo {author} {\bibfnamefont {A.~P.}\ \bibnamefont {Badia}},
  \bibinfo {author} {\bibfnamefont {M.}~\bibnamefont {Mirza}}, \bibinfo
  {author} {\bibfnamefont {A.}~\bibnamefont {Graves}}, \bibinfo {author}
  {\bibfnamefont {T.}~\bibnamefont {Lillicrap}}, \bibinfo {author}
  {\bibfnamefont {T.}~\bibnamefont {Harley}}, \bibinfo {author} {\bibfnamefont
  {D.}~\bibnamefont {Silver}},\ and\ \bibinfo {author} {\bibfnamefont
  {K.}~\bibnamefont {Kavukcuoglu}},\ }\bibfield  {title} {\bibinfo {title}
  {Asynchronous {Methods} for {Deep} {Reinforcement} {Learning}},\ }in\ \href
  {https://proceedings.mlr.press/v48/mniha16.html} {\emph {\bibinfo {booktitle}
  {Proceedings of {The} 33rd {International} {Conference} on {Machine}
  {Learning}}}},\ \bibinfo {series} {Proceedings of {Machine} {Learning}
  {Research}}, Vol.~\bibinfo {volume} {48},\ \bibinfo {editor} {edited by\
  \bibinfo {editor} {\bibfnamefont {M.~F.}\ \bibnamefont {Balcan}}\ and\
  \bibinfo {editor} {\bibfnamefont {K.~Q.}\ \bibnamefont {Weinberger}}}\
  (\bibinfo  {publisher} {PMLR},\ \bibinfo {address} {New York, New York,
  USA},\ \bibinfo {year} {2016})\ pp.\ \bibinfo {pages}
  {1928--1937}\BibitemShut {NoStop}%
\bibitem [{\citenamefont {Schulman}\ \emph {et~al.}(2017)\citenamefont
  {Schulman}, \citenamefont {Wolski}, \citenamefont {Dhariwal}, \citenamefont
  {Radford},\ and\ \citenamefont {Klimov}}]{schulman_proximal_2017}%
  \BibitemOpen
  \bibfield  {author} {\bibinfo {author} {\bibfnamefont {J.}~\bibnamefont
  {Schulman}}, \bibinfo {author} {\bibfnamefont {F.}~\bibnamefont {Wolski}},
  \bibinfo {author} {\bibfnamefont {P.}~\bibnamefont {Dhariwal}}, \bibinfo
  {author} {\bibfnamefont {A.}~\bibnamefont {Radford}},\ and\ \bibinfo {author}
  {\bibfnamefont {O.}~\bibnamefont {Klimov}},\ }\href
  {http://arxiv.org/abs/1707.06347} {\bibinfo {title} {Proximal {Policy}
  {Optimization} {Algorithms}}} (\bibinfo {year} {2017}),\ \bibinfo {note}
  {arXiv:1707.06347 [cs]}\BibitemShut {NoStop}%
\bibitem [{\citenamefont {Garnier}\ \emph {et~al.}(2021)\citenamefont
  {Garnier}, \citenamefont {Viquerat}, \citenamefont {Rabault}, \citenamefont
  {Larcher}, \citenamefont {Kuhnle},\ and\ \citenamefont
  {Hachem}}]{garnier_review_2021}%
  \BibitemOpen
  \bibfield  {author} {\bibinfo {author} {\bibfnamefont {P.}~\bibnamefont
  {Garnier}}, \bibinfo {author} {\bibfnamefont {J.}~\bibnamefont {Viquerat}},
  \bibinfo {author} {\bibfnamefont {J.}~\bibnamefont {Rabault}}, \bibinfo
  {author} {\bibfnamefont {A.}~\bibnamefont {Larcher}}, \bibinfo {author}
  {\bibfnamefont {A.}~\bibnamefont {Kuhnle}},\ and\ \bibinfo {author}
  {\bibfnamefont {E.}~\bibnamefont {Hachem}},\ }\bibfield  {title} {\bibinfo
  {title} {A review on deep reinforcement learning for fluid mechanics},\
  }\href {https://doi.org/10.1016/j.compfluid.2021.104973} {\bibfield
  {journal} {\bibinfo  {journal} {Computers \& Fluids}\ }\textbf {\bibinfo
  {volume} {225}},\ \bibinfo {pages} {104973} (\bibinfo {year}
  {2021})}\BibitemShut {NoStop}%
\bibitem [{\citenamefont {Rabault}\ \emph {et~al.}(2020)\citenamefont
  {Rabault}, \citenamefont {Ren}, \citenamefont {Zhang}, \citenamefont {Tang},\
  and\ \citenamefont {Xu}}]{rabault_deep_2020}%
  \BibitemOpen
  \bibfield  {author} {\bibinfo {author} {\bibfnamefont {J.}~\bibnamefont
  {Rabault}}, \bibinfo {author} {\bibfnamefont {F.}~\bibnamefont {Ren}},
  \bibinfo {author} {\bibfnamefont {W.}~\bibnamefont {Zhang}}, \bibinfo
  {author} {\bibfnamefont {H.}~\bibnamefont {Tang}},\ and\ \bibinfo {author}
  {\bibfnamefont {H.}~\bibnamefont {Xu}},\ }\bibfield  {title} {\bibinfo
  {title} {Deep reinforcement learning in fluid mechanics: {A} promising method
  for both active flow control and shape optimization},\ }\href
  {https://doi.org/10.1007/s42241-020-0028-y} {\bibfield  {journal} {\bibinfo
  {journal} {Journal of Hydrodynamics}\ }\textbf {\bibinfo {volume} {32}},\
  \bibinfo {pages} {234} (\bibinfo {year} {2020})}\BibitemShut {NoStop}%
\bibitem [{\citenamefont {Reddy}\ \emph {et~al.}(2018)\citenamefont {Reddy},
  \citenamefont {Wong-Ng}, \citenamefont {Celani}, \citenamefont {Sejnowski},\
  and\ \citenamefont {Vergassola}}]{reddy_glider_2018}%
  \BibitemOpen
  \bibfield  {author} {\bibinfo {author} {\bibfnamefont {G.}~\bibnamefont
  {Reddy}}, \bibinfo {author} {\bibfnamefont {J.}~\bibnamefont {Wong-Ng}},
  \bibinfo {author} {\bibfnamefont {A.}~\bibnamefont {Celani}}, \bibinfo
  {author} {\bibfnamefont {T.~J.}\ \bibnamefont {Sejnowski}},\ and\ \bibinfo
  {author} {\bibfnamefont {M.}~\bibnamefont {Vergassola}},\ }\bibfield  {title}
  {\bibinfo {title} {Glider soaring via reinforcement learning in the field},\
  }\href {https://doi.org/10.1038/s41586-018-0533-0} {\bibfield  {journal}
  {\bibinfo  {journal} {Nature}\ }\textbf {\bibinfo {volume} {562}},\ \bibinfo
  {pages} {236} (\bibinfo {year} {2018})}\BibitemShut {NoStop}%
\bibitem [{\citenamefont {Novati}\ \emph {et~al.}(2019)\citenamefont {Novati},
  \citenamefont {Mahadevan},\ and\ \citenamefont
  {Koumoutsakos}}]{novati_controlled_2019}%
  \BibitemOpen
  \bibfield  {author} {\bibinfo {author} {\bibfnamefont {G.}~\bibnamefont
  {Novati}}, \bibinfo {author} {\bibfnamefont {L.}~\bibnamefont {Mahadevan}},\
  and\ \bibinfo {author} {\bibfnamefont {P.}~\bibnamefont {Koumoutsakos}},\
  }\bibfield  {title} {\bibinfo {title} {Controlled gliding and perching
  through deep-reinforcement-learning},\ }\href
  {https://doi.org/10.1103/PhysRevFluids.4.093902} {\bibfield  {journal}
  {\bibinfo  {journal} {Phys. Rev. Fluids}\ }\textbf {\bibinfo {volume} {4}},\
  \bibinfo {pages} {093902} (\bibinfo {year} {2019})},\ \bibinfo {note}
  {publisher: American Physical Society}\BibitemShut {NoStop}%
\bibitem [{\citenamefont {Biferale}\ \emph {et~al.}(2019)\citenamefont
  {Biferale}, \citenamefont {Bonaccorso}, \citenamefont {Buzzicotti},
  \citenamefont {Clark Di~Leoni},\ and\ \citenamefont
  {Gustavsson}}]{biferale_zermelos_2019}%
  \BibitemOpen
  \bibfield  {author} {\bibinfo {author} {\bibfnamefont {L.}~\bibnamefont
  {Biferale}}, \bibinfo {author} {\bibfnamefont {F.}~\bibnamefont
  {Bonaccorso}}, \bibinfo {author} {\bibfnamefont {M.}~\bibnamefont
  {Buzzicotti}}, \bibinfo {author} {\bibfnamefont {P.}~\bibnamefont {Clark
  Di~Leoni}},\ and\ \bibinfo {author} {\bibfnamefont {K.}~\bibnamefont
  {Gustavsson}},\ }\bibfield  {title} {\bibinfo {title} {Zermelo’s problem:
  {Optimal} point-to-point navigation in {2D} turbulent flows using
  reinforcement learning},\ }\href {https://doi.org/10.1063/1.5120370}
  {\bibfield  {journal} {\bibinfo  {journal} {Chaos: An Interdisciplinary
  Journal of Nonlinear Science}\ }\textbf {\bibinfo {volume} {29}},\ \bibinfo
  {pages} {103138} (\bibinfo {year} {2019})}\BibitemShut {NoStop}%
\bibitem [{\citenamefont {Alageshan}\ \emph {et~al.}(2020)\citenamefont
  {Alageshan}, \citenamefont {Verma}, \citenamefont {Bec},\ and\ \citenamefont
  {Pandit}}]{alageshan_machine_2020}%
  \BibitemOpen
  \bibfield  {author} {\bibinfo {author} {\bibfnamefont {J.~K.}\ \bibnamefont
  {Alageshan}}, \bibinfo {author} {\bibfnamefont {A.~K.}\ \bibnamefont
  {Verma}}, \bibinfo {author} {\bibfnamefont {J.}~\bibnamefont {Bec}},\ and\
  \bibinfo {author} {\bibfnamefont {R.}~\bibnamefont {Pandit}},\ }\bibfield
  {title} {\bibinfo {title} {Machine learning strategies for path-planning
  microswimmers in turbulent flows},\ }\href
  {https://doi.org/10.1103/PhysRevE.101.043110} {\bibfield  {journal} {\bibinfo
   {journal} {Physical Review E}\ }\textbf {\bibinfo {volume} {101}},\ \bibinfo
  {pages} {043110} (\bibinfo {year} {2020})},\ \bibinfo {note}
  {arXiv:1910.01728 [physics, stat]}\BibitemShut {NoStop}%
\bibitem [{\citenamefont {Gunnarson}\ \emph {et~al.}(2021)\citenamefont
  {Gunnarson}, \citenamefont {Mandralis}, \citenamefont {Novati}, \citenamefont
  {Koumoutsakos},\ and\ \citenamefont {Dabiri}}]{gunnarson_learning_2021}%
  \BibitemOpen
  \bibfield  {author} {\bibinfo {author} {\bibfnamefont {P.}~\bibnamefont
  {Gunnarson}}, \bibinfo {author} {\bibfnamefont {I.}~\bibnamefont
  {Mandralis}}, \bibinfo {author} {\bibfnamefont {G.}~\bibnamefont {Novati}},
  \bibinfo {author} {\bibfnamefont {P.}~\bibnamefont {Koumoutsakos}},\ and\
  \bibinfo {author} {\bibfnamefont {J.~O.}\ \bibnamefont {Dabiri}},\ }\bibfield
   {title} {\bibinfo {title} {Learning efficient navigation in vortical flow
  fields},\ }\href {https://doi.org/10.1038/s41467-021-27015-y} {\bibfield
  {journal} {\bibinfo  {journal} {Nature Communications}\ }\textbf {\bibinfo
  {volume} {12}},\ \bibinfo {pages} {7143} (\bibinfo {year}
  {2021})}\BibitemShut {NoStop}%
\bibitem [{\citenamefont {Borra}\ \emph {et~al.}(2022)\citenamefont {Borra},
  \citenamefont {Biferale}, \citenamefont {Cencini},\ and\ \citenamefont
  {Celani}}]{borra_reinforcement_2022}%
  \BibitemOpen
  \bibfield  {author} {\bibinfo {author} {\bibfnamefont {F.}~\bibnamefont
  {Borra}}, \bibinfo {author} {\bibfnamefont {L.}~\bibnamefont {Biferale}},
  \bibinfo {author} {\bibfnamefont {M.}~\bibnamefont {Cencini}},\ and\ \bibinfo
  {author} {\bibfnamefont {A.}~\bibnamefont {Celani}},\ }\bibfield  {title}
  {\bibinfo {title} {Reinforcement learning for pursuit and evasion of
  microswimmers at low {Reynolds} number},\ }\href
  {https://doi.org/10.1103/PhysRevFluids.7.023103} {\bibfield  {journal}
  {\bibinfo  {journal} {Physical Review Fluids}\ }\textbf {\bibinfo {volume}
  {7}},\ \bibinfo {pages} {023103} (\bibinfo {year} {2022})}\BibitemShut
  {NoStop}%
\bibitem [{\citenamefont {Calascibetta}\ \emph
  {et~al.}(2023{\natexlab{a}})\citenamefont {Calascibetta}, \citenamefont
  {Biferale}, \citenamefont {Borra}, \citenamefont {Celani},\ and\
  \citenamefont {Cencini}}]{calascibetta_taming_2023}%
  \BibitemOpen
  \bibfield  {author} {\bibinfo {author} {\bibfnamefont {C.}~\bibnamefont
  {Calascibetta}}, \bibinfo {author} {\bibfnamefont {L.}~\bibnamefont
  {Biferale}}, \bibinfo {author} {\bibfnamefont {F.}~\bibnamefont {Borra}},
  \bibinfo {author} {\bibfnamefont {A.}~\bibnamefont {Celani}},\ and\ \bibinfo
  {author} {\bibfnamefont {M.}~\bibnamefont {Cencini}},\ }\bibfield  {title}
  {\bibinfo {title} {Taming {Lagrangian} chaos with multi-objective
  reinforcement learning},\ }\href
  {https://doi.org/10.1140/epje/s10189-023-00271-0} {\bibfield  {journal}
  {\bibinfo  {journal} {The European Physical Journal E}\ }\textbf {\bibinfo
  {volume} {46}},\ \bibinfo {pages} {9} (\bibinfo {year}
  {2023}{\natexlab{a}})}\BibitemShut {NoStop}%
\bibitem [{\citenamefont {Calascibetta}\ \emph
  {et~al.}(2023{\natexlab{b}})\citenamefont {Calascibetta}, \citenamefont
  {Biferale}, \citenamefont {Borra}, \citenamefont {Celani},\ and\
  \citenamefont {Cencini}}]{calascibetta_optimal_2023}%
  \BibitemOpen
  \bibfield  {author} {\bibinfo {author} {\bibfnamefont {C.}~\bibnamefont
  {Calascibetta}}, \bibinfo {author} {\bibfnamefont {L.}~\bibnamefont
  {Biferale}}, \bibinfo {author} {\bibfnamefont {F.}~\bibnamefont {Borra}},
  \bibinfo {author} {\bibfnamefont {A.}~\bibnamefont {Celani}},\ and\ \bibinfo
  {author} {\bibfnamefont {M.}~\bibnamefont {Cencini}},\ }\bibfield  {title}
  {\bibinfo {title} {Optimal tracking strategies in a turbulent flow},\ }\href
  {https://doi.org/10.1038/s42005-023-01366-y} {\bibfield  {journal} {\bibinfo
  {journal} {Communications Physics}\ }\textbf {\bibinfo {volume} {6}},\
  \bibinfo {pages} {256} (\bibinfo {year} {2023}{\natexlab{b}})}\BibitemShut
  {NoStop}%
\bibitem [{\citenamefont {Borra}\ \emph {et~al.}(2021)\citenamefont {Borra},
  \citenamefont {Cencini},\ and\ \citenamefont {Celani}}]{borra_optimal_2021}%
  \BibitemOpen
  \bibfield  {author} {\bibinfo {author} {\bibfnamefont {F.}~\bibnamefont
  {Borra}}, \bibinfo {author} {\bibfnamefont {M.}~\bibnamefont {Cencini}},\
  and\ \bibinfo {author} {\bibfnamefont {A.}~\bibnamefont {Celani}},\
  }\bibfield  {title} {\bibinfo {title} {Optimal collision avoidance in swarms
  of active {Brownian} particles},\ }\href
  {https://doi.org/10.1088/1742-5468/ac12c6} {\bibfield  {journal} {\bibinfo
  {journal} {Journal of Statistical Mechanics: Theory and Experiment}\ }\textbf
  {\bibinfo {volume} {2021}},\ \bibinfo {pages} {083401} (\bibinfo {year}
  {2021})}\BibitemShut {NoStop}%
\bibitem [{\citenamefont {van Hasselt}\ \emph {et~al.}(2016)\citenamefont {van
  Hasselt}, \citenamefont {Guez}, \citenamefont {Hessel}, \citenamefont
  {Mnih},\ and\ \citenamefont {Silver}}]{acrossvalues}%
  \BibitemOpen
  \bibfield  {author} {\bibinfo {author} {\bibfnamefont {H.}~\bibnamefont {van
  Hasselt}}, \bibinfo {author} {\bibfnamefont {A.}~\bibnamefont {Guez}},
  \bibinfo {author} {\bibfnamefont {M.}~\bibnamefont {Hessel}}, \bibinfo
  {author} {\bibfnamefont {V.}~\bibnamefont {Mnih}},\ and\ \bibinfo {author}
  {\bibfnamefont {D.}~\bibnamefont {Silver}},\ }\href@noop {} {\bibinfo {title}
  {Learning values across many orders of magnitude}} (\bibinfo {year} {2016}),\
  \Eprint {https://arxiv.org/abs/1602.07714} {arXiv:1602.07714 [cs.LG]}
  \BibitemShut {NoStop}%
\bibitem [{\citenamefont {Uchendu}\ \emph {et~al.}(2023)\citenamefont
  {Uchendu}, \citenamefont {Xiao}, \citenamefont {Lu}, \citenamefont {Zhu},
  \citenamefont {Yan}, \citenamefont {Simon}, \citenamefont {Bennice},
  \citenamefont {Fu}, \citenamefont {Ma}, \citenamefont {Jiao}, \citenamefont
  {Levine},\ and\ \citenamefont {Hausman}}]{uchendu2023jumpstart}%
  \BibitemOpen
  \bibfield  {author} {\bibinfo {author} {\bibfnamefont {I.}~\bibnamefont
  {Uchendu}}, \bibinfo {author} {\bibfnamefont {T.}~\bibnamefont {Xiao}},
  \bibinfo {author} {\bibfnamefont {Y.}~\bibnamefont {Lu}}, \bibinfo {author}
  {\bibfnamefont {B.}~\bibnamefont {Zhu}}, \bibinfo {author} {\bibfnamefont
  {M.}~\bibnamefont {Yan}}, \bibinfo {author} {\bibfnamefont {J.}~\bibnamefont
  {Simon}}, \bibinfo {author} {\bibfnamefont {M.}~\bibnamefont {Bennice}},
  \bibinfo {author} {\bibfnamefont {C.}~\bibnamefont {Fu}}, \bibinfo {author}
  {\bibfnamefont {C.}~\bibnamefont {Ma}}, \bibinfo {author} {\bibfnamefont
  {J.}~\bibnamefont {Jiao}}, \bibinfo {author} {\bibfnamefont {S.}~\bibnamefont
  {Levine}},\ and\ \bibinfo {author} {\bibfnamefont {K.}~\bibnamefont
  {Hausman}},\ }\href@noop {} {\bibinfo {title} {Jump-start reinforcement
  learning}} (\bibinfo {year} {2023}),\ \Eprint
  {https://arxiv.org/abs/2204.02372} {arXiv:2204.02372 [cs.LG]} \BibitemShut
  {NoStop}%
\bibitem [{\citenamefont {Mnih}(2013)}]{mnih2013playing}%
  \BibitemOpen
  \bibfield  {author} {\bibinfo {author} {\bibfnamefont {V.}~\bibnamefont
  {Mnih}},\ }\bibfield  {title} {\bibinfo {title} {Playing atari with deep
  reinforcement learning},\ }\href@noop {} {\bibfield  {journal} {\bibinfo
  {journal} {arXiv preprint arXiv:1312.5602}\ } (\bibinfo {year}
  {2013})}\BibitemShut {NoStop}%
\bibitem [{\citenamefont {Batchelor}(1959)}]{batchelor_small-scale_1959}%
  \BibitemOpen
  \bibfield  {author} {\bibinfo {author} {\bibfnamefont {G.~K.}\ \bibnamefont
  {Batchelor}},\ }\bibfield  {title} {\bibinfo {title} {Small-scale variation
  of convected quantities like temperature in turbulent fluid {Part} 1.
  {General} discussion and the case of small conductivity},\ }\href
  {https://doi.org/10.1017/S002211205900009X} {\bibfield  {journal} {\bibinfo
  {journal} {Journal of Fluid Mechanics}\ }\textbf {\bibinfo {volume} {5}},\
  \bibinfo {pages} {113} (\bibinfo {year} {1959})},\ \bibinfo {note}
  {publisher: Cambridge University Press}\BibitemShut {NoStop}%
\bibitem [{\citenamefont {Kraichnan}(1968)}]{kraichnan_small-scale_1968}%
  \BibitemOpen
  \bibfield  {author} {\bibinfo {author} {\bibfnamefont {R.~H.}\ \bibnamefont
  {Kraichnan}},\ }\bibfield  {title} {\bibinfo {title} {Small-{Scale}
  {Structure} of a {Scalar} {Field} {Convected} by {Turbulence}},\ }\href
  {https://doi.org/10.1063/1.1692063} {\bibfield  {journal} {\bibinfo
  {journal} {Physics of Fluids}\ }\textbf {\bibinfo {volume} {11}},\ \bibinfo
  {pages} {945} (\bibinfo {year} {1968})}\BibitemShut {NoStop}%
\bibitem [{\citenamefont {Shraiman}\ and\ \citenamefont
  {Siggia}(1994)}]{shraiman_lagrangian_1994}%
  \BibitemOpen
  \bibfield  {author} {\bibinfo {author} {\bibfnamefont {B.~I.}\ \bibnamefont
  {Shraiman}}\ and\ \bibinfo {author} {\bibfnamefont {E.~D.}\ \bibnamefont
  {Siggia}},\ }\bibfield  {title} {\bibinfo {title} {Lagrangian path integrals
  and fluctuations in random flow},\ }\href
  {https://doi.org/10.1103/PhysRevE.49.2912} {\bibfield  {journal} {\bibinfo
  {journal} {Physical Review E}\ }\textbf {\bibinfo {volume} {49}},\ \bibinfo
  {pages} {2912} (\bibinfo {year} {1994})}\BibitemShut {NoStop}%
\bibitem [{\citenamefont {Chertkov}\ \emph {et~al.}(1995)\citenamefont
  {Chertkov}, \citenamefont {Falkovich}, \citenamefont {Kolokolov},\ and\
  \citenamefont {Lebedev}}]{chertkov_statistics_1995}%
  \BibitemOpen
  \bibfield  {author} {\bibinfo {author} {\bibfnamefont {M.}~\bibnamefont
  {Chertkov}}, \bibinfo {author} {\bibfnamefont {G.}~\bibnamefont {Falkovich}},
  \bibinfo {author} {\bibfnamefont {I.}~\bibnamefont {Kolokolov}},\ and\
  \bibinfo {author} {\bibfnamefont {V.}~\bibnamefont {Lebedev}},\ }\bibfield
  {title} {\bibinfo {title} {Statistics of a passive scalar advected by a
  large-scale two-dimensional velocity field: {Analytic} solution},\ }\href
  {https://doi.org/10.1103/PhysRevE.51.5609} {\bibfield  {journal} {\bibinfo
  {journal} {Phys. Rev. E}\ }\textbf {\bibinfo {volume} {51}},\ \bibinfo
  {pages} {5609} (\bibinfo {year} {1995})},\ \bibinfo {note} {publisher:
  American Physical Society}\BibitemShut {NoStop}%
\bibitem [{\citenamefont {Bernard}\ \emph {et~al.}(1998)\citenamefont
  {Bernard}, \citenamefont {Gawedzki},\ and\ \citenamefont
  {Kupiainen}}]{bernard_slow_1998}%
  \BibitemOpen
  \bibfield  {author} {\bibinfo {author} {\bibfnamefont {D.}~\bibnamefont
  {Bernard}}, \bibinfo {author} {\bibfnamefont {K.}~\bibnamefont {Gawedzki}},\
  and\ \bibinfo {author} {\bibfnamefont {A.}~\bibnamefont {Kupiainen}},\
  }\bibfield  {title} {\bibinfo {title} {Slow {Modes} in {Passive}
  {Advection}},\ }\href {https://doi.org/10.1023/A:1023212600779} {\bibfield
  {journal} {\bibinfo  {journal} {Journal of Statistical Physics}\ }\textbf
  {\bibinfo {volume} {90}},\ \bibinfo {pages} {519} (\bibinfo {year}
  {1998})}\BibitemShut {NoStop}%
\bibitem [{\citenamefont {Balkovsky}\ and\ \citenamefont
  {Fouxon}(1999)}]{balkovsky_universal_1999}%
  \BibitemOpen
  \bibfield  {author} {\bibinfo {author} {\bibfnamefont {E.}~\bibnamefont
  {Balkovsky}}\ and\ \bibinfo {author} {\bibfnamefont {A.}~\bibnamefont
  {Fouxon}},\ }\bibfield  {title} {\bibinfo {title} {Universal long-time
  properties of {Lagrangian} statistics in the {Batchelor} regime and their
  application to the passive scalar problem},\ }\href
  {https://doi.org/10.1103/PhysRevE.60.4164} {\bibfield  {journal} {\bibinfo
  {journal} {Phys. Rev. E}\ }\textbf {\bibinfo {volume} {60}},\ \bibinfo
  {pages} {4164} (\bibinfo {year} {1999})},\ \bibinfo {note} {publisher:
  American Physical Society}\BibitemShut {NoStop}%
\bibitem [{\citenamefont {Falkovich}\ \emph {et~al.}(2001)\citenamefont
  {Falkovich}, \citenamefont {Gawedzki},\ and\ \citenamefont
  {Vergassola}}]{falkovich_particles_2001}%
  \BibitemOpen
  \bibfield  {author} {\bibinfo {author} {\bibfnamefont {G.}~\bibnamefont
  {Falkovich}}, \bibinfo {author} {\bibfnamefont {K.}~\bibnamefont
  {Gawedzki}},\ and\ \bibinfo {author} {\bibfnamefont {M.}~\bibnamefont
  {Vergassola}},\ }\bibfield  {title} {\bibinfo {title} {Particles and fields
  in fluid turbulence},\ }\href {https://doi.org/10.1103/RevModPhys.73.913}
  {\bibfield  {journal} {\bibinfo  {journal} {Rev. Mod. Phys.}\ }\textbf
  {\bibinfo {volume} {73}},\ \bibinfo {pages} {913} (\bibinfo {year} {2001})},\
  \bibinfo {note} {publisher: American Physical Society}\BibitemShut {NoStop}%
\bibitem [{\citenamefont {Ruelle}(1979)}]{ruelle_ergodic_1979}%
  \BibitemOpen
  \bibfield  {author} {\bibinfo {author} {\bibfnamefont {D.}~\bibnamefont
  {Ruelle}},\ }\bibfield  {title} {\bibinfo {title} {Ergodic theory of
  differentiable dynamical systems},\ }\href
  {https://doi.org/10.1007/BF02684768} {\bibfield  {journal} {\bibinfo
  {journal} {Publications mathématiques de l'IHÉS}\ }\textbf {\bibinfo
  {volume} {50}},\ \bibinfo {pages} {27} (\bibinfo {year} {1979})}\BibitemShut
  {NoStop}%
\bibitem [{\citenamefont {Goldhirsch}\ \emph {et~al.}(1987)\citenamefont
  {Goldhirsch}, \citenamefont {Sulem},\ and\ \citenamefont
  {Orszag}}]{goldhirsch_stability_1987}%
  \BibitemOpen
  \bibfield  {author} {\bibinfo {author} {\bibfnamefont {I.}~\bibnamefont
  {Goldhirsch}}, \bibinfo {author} {\bibfnamefont {P.-L.}\ \bibnamefont
  {Sulem}},\ and\ \bibinfo {author} {\bibfnamefont {S.~A.}\ \bibnamefont
  {Orszag}},\ }\bibfield  {title} {\bibinfo {title} {Stability and {Lyapunov}
  stability of dynamical systems: {A} differential approach and a numerical
  method},\ }\href {https://doi.org/10.1016/0167-2789(87)90034-0} {\bibfield
  {journal} {\bibinfo  {journal} {Physica D: Nonlinear Phenomena}\ }\textbf
  {\bibinfo {volume} {27}},\ \bibinfo {pages} {311} (\bibinfo {year}
  {1987})}\BibitemShut {NoStop}%
\bibitem [{\citenamefont {Chertkov}\ \emph {et~al.}(1994)\citenamefont
  {Chertkov}, \citenamefont {Gamba},\ and\ \citenamefont
  {Kolokolov}}]{chertkov_exact_1994}%
  \BibitemOpen
  \bibfield  {author} {\bibinfo {author} {\bibfnamefont {M.}~\bibnamefont
  {Chertkov}}, \bibinfo {author} {\bibfnamefont {A.}~\bibnamefont {Gamba}},\
  and\ \bibinfo {author} {\bibfnamefont {I.}~\bibnamefont {Kolokolov}},\
  }\bibfield  {title} {\bibinfo {title} {Exact field-theoretical description of
  passive scalar convection in an {N}-dimensional long-range velocity field},\
  }\href {https://doi.org/10.1016/0375-9601(94)90233-X} {\bibfield  {journal}
  {\bibinfo  {journal} {Physics Letters A}\ }\textbf {\bibinfo {volume}
  {192}},\ \bibinfo {pages} {435} (\bibinfo {year} {1994})}\BibitemShut
  {NoStop}%
\bibitem [{\citenamefont {Chertkov}(1998)}]{chertkov_how_1998}%
  \BibitemOpen
  \bibfield  {author} {\bibinfo {author} {\bibfnamefont {M.}~\bibnamefont
  {Chertkov}},\ }\bibfield  {title} {\bibinfo {title} {On how a joint
  interaction of two innocent partners (smooth advection and linear damping)
  produces a strong intermittency},\ }\href {https://doi.org/10.1063/1.869826}
  {\bibfield  {journal} {\bibinfo  {journal} {Physics of Fluids}\ }\textbf
  {\bibinfo {volume} {10}},\ \bibinfo {pages} {3017} (\bibinfo {year}
  {1998})}\BibitemShut {NoStop}%
\bibitem [{\citenamefont {Zhao}\ \emph {et~al.}(1993)\citenamefont {Zhao},
  \citenamefont {Kwek}, \citenamefont {Li},\ and\ \citenamefont
  {Huang}}]{zhao_chaotic_1993}%
  \BibitemOpen
  \bibfield  {author} {\bibinfo {author} {\bibfnamefont {X.-H.}\ \bibnamefont
  {Zhao}}, \bibinfo {author} {\bibfnamefont {K.-H.}\ \bibnamefont {Kwek}},
  \bibinfo {author} {\bibfnamefont {J.-B.}\ \bibnamefont {Li}},\ and\ \bibinfo
  {author} {\bibfnamefont {K.-L.}\ \bibnamefont {Huang}},\ }\bibfield  {title}
  {\bibinfo {title} {Chaotic and {Resonant} {Streamlines} in the {ABC}
  {Flow}},\ }\href {http://www.jstor.org/stable/2102274} {\bibfield  {journal}
  {\bibinfo  {journal} {SIAM Journal on Applied Mathematics}\ }\textbf
  {\bibinfo {volume} {53}},\ \bibinfo {pages} {71} (\bibinfo {year} {1993})},\
  \bibinfo {note} {publisher: Society for Industrial and Applied
  Mathematics}\BibitemShut {NoStop}%
\bibitem [{\citenamefont {Novati}\ \emph {et~al.}(2021)\citenamefont {Novati},
  \citenamefont {De~Laroussilhe},\ and\ \citenamefont
  {Koumoutsakos}}]{novati_automating_2021}%
  \BibitemOpen
  \bibfield  {author} {\bibinfo {author} {\bibfnamefont {G.}~\bibnamefont
  {Novati}}, \bibinfo {author} {\bibfnamefont {H.~L.}\ \bibnamefont
  {De~Laroussilhe}},\ and\ \bibinfo {author} {\bibfnamefont {P.}~\bibnamefont
  {Koumoutsakos}},\ }\bibfield  {title} {\bibinfo {title} {Automating
  turbulence modelling by multi-agent reinforcement learning},\ }\href
  {https://doi.org/10.1038/s42256-020-00272-0} {\bibfield  {journal} {\bibinfo
  {journal} {Nature Machine Intelligence}\ }\textbf {\bibinfo {volume} {3}},\
  \bibinfo {pages} {87} (\bibinfo {year} {2021})}\BibitemShut {NoStop}%
\bibitem [{\citenamefont {Bae}\ and\ \citenamefont
  {Koumoutsakos}(2022)}]{bae_scientific_2022}%
  \BibitemOpen
  \bibfield  {author} {\bibinfo {author} {\bibfnamefont {H.~J.}\ \bibnamefont
  {Bae}}\ and\ \bibinfo {author} {\bibfnamefont {P.}~\bibnamefont
  {Koumoutsakos}},\ }\bibfield  {title} {\bibinfo {title} {Scientific
  multi-agent reinforcement learning for wall-models of turbulent flows},\
  }\href {https://doi.org/10.1038/s41467-022-28957-7} {\bibfield  {journal}
  {\bibinfo  {journal} {Nature Communications}\ }\textbf {\bibinfo {volume}
  {13}},\ \bibinfo {pages} {1443} (\bibinfo {year} {2022})}\BibitemShut
  {NoStop}%
\end{thebibliography}%

\end{document}